\documentclass[usenatbib]{jpp}
\usepackage[applemac]{inputenc}
\usepackage{graphicx}
\usepackage{amssymb}
\usepackage{epstopdf}
\usepackage[authoryear]{natbib}
\usepackage{amsmath, float, bm}
\usepackage{txfonts}
\usepackage{fleqn}   
\usepackage{amssymb}
\usepackage{enumerate}
\usepackage{graphicx, subfigure}
\usepackage{calligra}
\usepackage{lscape}
\usepackage{tabularx}
\usepackage{setspace}
\usepackage[dvipsnames]{xcolor}
\usepackage{epstopdf, epsfig}
\usepackage{csquotes}
\usepackage{color}

\def\degr{%
    \relax
    \ifmmode  ^\circ
    \else  $^\circ$%
    \fi
}
\def\arcsec{\hbox{$^{\prime\prime}$}}

\shorttitle{Multi-D MHD modelling of PWNe}
\shortauthor{B.~Olmi, L.~Del Zanna, E.~Amato, N.~Bucciantini, A.~Mignone}

\title{Multi-D magnetohydrodynamic modelling of pulsar wind nebulae: recent progress and open questions}

\author{B.~Olmi\aff{1,2,3}
  \corresp{\email{barbara.olmi@unifi.it}},
  L.~Del~Zanna\aff{1,2,3},
  E.~Amato\aff{2,1},
  N.~Bucciantini\aff{2,1,3},
  A.~Mignone\aff{4}}

\affiliation{\aff{1} Dipartimento di Fisica ed Astronomia, Università degli Studi di Firenze, Via G. Sansone 1, 50019 Sesto F.no (Firenze), Italy
	\aff{2} INAF Osservatorio Astrofisico di Arcetri, Largo E. Fermi 5, 50125 Firenze, Italy
	\aff{3} INFN Sezione di Firenze, Via G. Sansone 1, 50019 Sesto F.no (Firenze), Italy
	\aff{4} Dipartimento di Fisica Generale \enquote{Amedeo Avogadro} Università degli Studi di Torino, Via Pietro Giuria 1, 10125 Torino, Italy
	}

\begin{document}

\maketitle

\begin{abstract}
In the last decade, the relativistic magnetohydrodynamic (MHD) modelling of pulsar wind nebulae, and of the Crab nebula in particular, has been highly successful, with many of the observed dynamical and emission properties reproduced down to the finest detail. 
Here, we critically discuss the results of some of the most recent studies: namely the investigation of the origin of the radio emitting particles and the quest for the acceleration sites of particles of different energies along the termination shock, by using wisps motion as a diagnostic tool; the study of the magnetic dissipation process in high magnetization nebulae by means of new long-term three-dimensional simulations of the pulsar wind nebula evolution; the investigation of the relativistic tearing instability in thinning current sheets, leading to fast reconnection events that might be at the origin of the Crab nebula gamma-ray flares.
\end{abstract}

\section{Introduction}
Pulsar Wind Nebulae (PWNe) represent the best laboratory for relativistic plasma astrophysics and high-energy astrophysics, since the very extreme conditions which can be found in these objects cannot be obtained artificially in a laboratory. 
They are a particular class of supernova remnants, also known as \textit{plerions} due to their centre filled emission, which makes them easily distinguishable from shell-type remnants, which on the contrary show a limb-brightened morphology.

At present, approximately 100 PWNe are known, and they all show a broad band spectrum, extending from radio to X-ray or even gamma-ray frequencies \citep{Gaensler:2006, Kargaltsev:2013}. 
Spectra are dominated by non-thermal synchrotron emission, due to relativistic particles moving in the nebular magnetic field, from radio up to a few MeV, and by inverse Compton scattering (IC) at higher energies. The radio component is characterized by a flat spectral index, typically in the range $-0.3 \lesssim \alpha_r \lesssim 0$, while the high-energy component has a much steeper spectrum, with typical photon indices $\Gamma \sim 1{\rm -}2$: this implies the presence of one or more spectral breaks between the radio and X-ray frequencies \citep{Gaensler:2006}.

PWNe are basically bubbles of relativistic particles and electromagnetic fields, born in the core collapse supernova explosion of massive stars ($M\gtrsim 8 M_\odot$). The nucleus of the progenitor star collapses to a highly magnetized and rapidly rotating neutron star, often observed as a pulsar.  
When the supernova event terminates, the majority of the rotational energy lost by the neutron star fills the remnant with a magnetized, cold, relativistic plasma wind, mainly made of electron-positron pairs and possibly a small amount of hadrons (ions) \citep{Gallant:1994}. 
The pairs are copiously produced in the pulsar magnetosphere where the primary electrons, extracted from the star surface by the extremely strong induced electric field, interact with radiation and give rise to an electromagnetic cascade. The exact amount of pair production in the pulsar magnetosphere and its dependence on the pulsar parameters is one of the missing ingredients in our understanding of pulsar physics and a subject of active investigation \citep{Hibschman:2001,Timokhin:2015}. 
Constraints on this unknown quantity deriving from PWNe modelling will be extensively discussed in the section \ref{sec:radioO}.

In the case of a young pulsar (usually Crab-like PWNe are observed for pulsars younger than $\sim 20000$ yrs), the emitted wind is still surrounded by the denser and slower debris of the supernova explosion, which are expanding outward with non-relativistic velocity. The interaction with this material generates a reverse shock, that propagates back towards the pulsar. If the magnetic energy in the wind does not exceed by far the particle kinetic energy, then the reverse shock will propagate back only down to a distance where the wind ram pressure upstream equals the nebular pressure downstream. From this point on, this termination shock (TS hereafter) will slowly move outward, as the nebula expands, as long as the pulsar input can be considered constant in time. At the TS, the pulsar wind is slowed down and heated: its bulk energy is efficiently dissipated and it goes in accelerated particles, which will be responsible for the observed nebular synchrotron emission across the electromagnetic spectrum \citep{Pacini:1973, Rees:1974, Kennel:1984}. Between $\sim 10\%$-30\% of the energy lost by the pulsar shows up as synchrotron emission of the daughter PWN \citep{Kennel:1984a}, to be compared with the ten times smaller fraction that typically appears as gamma-ray pulsed emission (and in radio it is even much smaller). It is appropriate to notice that in the case of a strongly magnetized wind (Poynting flux much larger than the particle kinetic energy flux) this picture must be modified somewhat: a reverse shock hitting such a wind will likely produce little dissipation and possibly never reach a self-similar stage of expansion. In any case in PWNe effective dissipation is usually inferred to take place in a very thin layer of plasma, which either means a strong magnetohydrodynamic (MHD) shock develops in a moderately magnetized medium or some non-MHD effects (likely magnetic reconnection) guarantee the same result.

Thanks to its proximity ($\sim 2$ kpc) and to the high emitted power, the Crab nebula is unique among PWNe, and it is naturally considered as the prototype of the whole class. It is indeed one of the best studied objects in the sky, and the one from which we have learnt most of what we currently know about PWNe \citep{Hester:2008}. The Crab nebula is almost certainly associated with a supernova explosion that happened in 1054~AD. Chinese astronomers reported that a \enquote{guest} star appeared in the sky and that it was visible during daytime for three weeks and for 22 months at night \citep[and references therein]{Stephenson:2002}. The nebular remnant was then discovered in 1731 by John Bevis, and observed a few years later by Charles Messier, becoming the first item of his famous catalogue of non-cometary objects, with the name M1.

As inferred from energetic arguments by \citet{Pacini:1967} prior to the actual discovery of pulsars, the central engine of the Crab nebula was indeed identified as a pulsar rotating with period $P=33$~ms, and slowly increasing with time ($\dot{P}=4.21\times 10^{-13}$ s s$^{-1}$). This observed spin-down implies that the rotational energy is dissipated at a rate of $L\sim 5 \times 10^{38}$ erg s$^{-1}$, equivalent to $1.3\times 10^{5}~L_\odot$, a value similar to the inferred rate at which the energy is supplied to the nebula \citep{Gold:1969}.

After the launch of the Chandra X-ray observatory, the high-energy morphology of the Crab nebula and of many other PWNe became accessible, and many different variable and bright features were observed in the inner regions with great detail. A puzzling and unexpected jet-torus structure was identified in many objects, including the Crab nebula \citep{Hester:1995, Helfand:2001, Pavlov:2003, Gaensler:2001, Gaensler:2002, Lu:2002}. The Crab torus, in particular, appeared to be composed of several rings, the brightest of which is located at a distance of $\sim 0.1$~pc from the pulsar. This ring, usually referred to as \enquote{inner ring}, is likely associated with the TS location \citep{Weisskopf:2000}.

In the optical band, \citet{Scargle:1969} had previously identified many bright, arc-like, variable structures, that he named \enquote{wisps}. These are currently known to fluctuate in brightness on very short time scales (of the order of ks \citep{Hester:2002, Melatos:2005}). They are also seen to move outward with mildly relativistic velocities ($\lesssim0.5c$, with $c$ the speed of light), and progressively fade. The typical period between their appearance close to the inner ring and disappearance in the bulk of the nebula can be of the order of hours, days, months or even years \citep{Melatos:2005, Bietenholz:2004,Hester:2002}, depending on the wavelength of observation (infra-red (IR), X-rays, optical and radio respectively). 
As we will see in the following, since wisps originate so close to the supposed TS location, where the flow is highly turbulent and swirling, they represent the best features to investigate the physical conditions of the plasma and of the emitting particles near to the shock surface.

Other variable features, the so-called \enquote{knots}, had been later observed by \citet{van-den-Bergh:1989} as point-like structures, that always appear and fade out at the same locations. These are visible at basically all wavelengths. In the optical band the most luminous one (called knot-1) is located at $0.65\arcsec$ from the pulsar, while the second brightest one is at a distance of $3.8\arcsec$. All knots appear to be aligned with the polar jet in the south-east (SE) direction. 

The presence of polar X-ray jets is certainly one of the most intriguing features in the high-energy morphology of PWNe. They appear to originate so close to the pulsar location that, when first discovered, it was natural to think that any collimation mechanism had to operate upstream of the TS, in the pulsar wind region. The best candidate among the possible models was soon recognized to be collimation by the hoop stresses due to the dominant toroidal magnetic field component in the wind, but this can be shown to be very inefficient inside an ultra-relativistic MHD flow, such as the pulsar wind before the TS. Later \citet{Lyubarsky:2002} suggested that an anisotropic distribution of the energy flux in the wind could cause the shock to be highly non-spherical, rather developing a characteristic oblate shape \citep[see also][]{Bogovalov:2002}. 
The TS is more expanded in the equatorial direction, where the energy flux is higher, but is much closer to the pulsar along the polar axis. Upstream collimation of the flow is then only apparent, while in reality it is the downstream flow that is diverted towards the axis by hoop stresses and forms a jet. Thus, the collimation mechanism is actually operating in the downstream PWN plasma, not in the pulsar wind, and theoretical difficulties are avoided.

This scenario was soon successfully confirmed by means of axisymmetric relativistic MHD numerical simulations of PWNe \citep{Komissarov:2004,Del-Zanna:2004,Bogovalov:2005}. MHD numerical modelling has progressively become more and more sophisticated, first with the introduction of increasingly more accurate emission diagnostics, and more recently with the upgrade to full three-dimensional (3-D) simulations. The questions that one has attempted to answer have also become deeper. The description of the most recent progresses and of the physical implications of the numerical findings will be the subject of the present paper.

\section{The big open problems in PWN physics}
The final purpose of the increasingly refined methods of analysis that have been developed to study the relativistic plasma dynamics in PWNe is to answer a number of old and new questions, the most important of which, in our view, are shortly reviewed in this section.
\subsection{The pulsar wind magnetization}
The fundamental parameter that characterizes the physical behaviour of the (cold) pulsar wind, and hence also the properties of the PWN arising from its interaction with the surrounding supernova remnant material, is the magnetization $\sigma$. This is defined as the ratio between the Poynting and the particle kinetic energy fluxes in the wind:
\begin{equation}\label{eq_sigma}
	\sigma=\frac{B^2}{4\pi n m_e \gamma^2 c^2}\,,
\end{equation}	
where $\gamma, \, B$ and $n$ are the Lorentz factor, the magnetic field and the rest-frame particle number density of the flow, and $m_e$ is the electron mass (here we consider a wind with kinetic energy flux dominated by pairs). This quantity is expected to vary with distance from the central pulsar, but also with latitude from the equatorial plane of the pulsar rotation.
According to theoretical models of pulsar magnetospheres, the wind is expected to be highly magnetized near the pulsar light cylinder, where clearly $\sigma \gg 1$. 
On the other hand, the very first attempts at modelling the Crab nebula within the framework of ideal MHD, showed that effective deceleration of the flow at the TS is only possible if its magnetization is not too high. In particular, steady-state analytical MHD models of the Crab nebula \citep{Rees:1974,Kennel:1984,BegelmanLi:1992} require a value of $\sigma\sim v_n/c$, with $v_n$ the expansion velocity of the PWN, always much less than one (which in the case of the Crab nebula is $\sigma \sim 3\times 10^{-3}$). 
Assuming that the downstream flow is well described by ideal MHD, strong magnetic dissipation is then needed between the light cylinder and the wind TS.

Since pulsars are oblique rotators, the wind has a complex structure: the current sheet that separates magnetic field lines of opposite polarities twists and tangles, cutting the equatorial plane along twin spirals. These oscillations take place within a region around the equator of angular extent equal to the inclination angle between the rotational and magnetic axes of the pulsar.
Neighbouring stripes of opposite polarities offer a perfect location for magnetic dissipation to occur, and hence $\sigma$ can be substantially decreased in that region \citep{Coroniti:1990}.

Nevertheless, in the case of the Crab nebula, the difference between the value of $\sigma$ expected at the light cylinder and that estimated at the TS in a model similar to that of Kennel \& Coroniti requires an efficiency of conversion of magnetic field energy into particle kinetic energy which appears too high to be explained by magnetic reconnection in the striped wind \citep{Lyubarsky:2001}. In the last decade, 2-D axisymmetric MHD simulations actually pointed out that at the TS the magnetization is likely higher than estimated based on steady-state 1-D MHD, and in particular, for the Crab PWN $\sigma \gtrsim 0.01$ is required at the TS in order to generate polar jets \citep{Del-Zanna:2006}. However, even in the framework of 2D MHD, the estimated values of $\sigma$ at the TS are still much smaller than unity, leaving the problem of efficient magnetic dissipation along the way to the shock.

Recently \citet{Porth:2014} presented the first 3-D numerical simulation of the Crab nebula, showing that values of the magnetization greater than unity can be worked out by just allowing for non-axisymmetric models, by increasing the dimensionality of the simulations. 
Downstream of the shock, the onset of kink instabilities leads to an efficient conversion of a large fraction of the toroidal magnetic field in the nebula into a poloidal component, reducing the magnetic tension, and also producing a larger amount of dissipation with respect to that observed in two dimensions. 
These simulations, however, have been run only for approximately $1/10$ of the Crab lifetime ($\simeq 1000$ yr), and end when the system is still far from the self-similar expansion phase, since the ratio between the nebula radius and the TS radius is not yet saturated.
In particular, it is not clear yet whether the average nebular magnetic field strength will be close to that inferred from observations ($\sim 200 \,\mu$G) at the final stage of the evolution, and if the synthetic non-thermal emission will match observational data in all spectral bands, thus confirming the model assumptions on the pulsar wind and emitting particles properties.  
Further investigation is certainly needed in this respect.

\subsection{Origin of radio emitting particles and implications for pulsar multiplicity}\label{sec:radioO}
Although radio emitting particles are dominant by number in the Crab nebula, their origin is still a mystery.
Since pulsars are believed to be primary antimatter factories in the Galaxy, reasonably good knowledge of the amount of particles generated in their magnetospheres has fundamental implications, not only for pulsar physics, but also for quantifying their contribution to the measured cosmic ray positron excess \citep{PAMELA-coll.:2009,AMS-02-coll.:2013}, and to asses the role of other possible sources, as {\it e.g.} dark matter annihilation. 
The pair production in pulsar magnetospheres is still not completely understood, but there is a general consensus that each primary electron extracted from the star surface generates a high number $\kappa$ of pairs (the so called \enquote{pulsar pair multiplicity}) in the cascading process. For young, energetic pulsars, the expected value of the multiplicity is in the range $10^3\lesssim \kappa \lesssim 10^7$.

Since secondary particles become part of the pulsar wind, a direct estimate of $\kappa$ can be obtained by accurate radiation modelling of the nebula.
An important thing to notice is that ions can only be primary, i.e. extracted from the star surface, since they cannot be generated in electromagnetic cascades. As a consequence they can only be at most a fraction $\sim 1/\kappa$ of the total number of the wind pairs. Nevertheless, if $\kappa<m_i/m_e\approx 10^3-10^4$, with $m_i$ the ion mass, the hadronic component of the wind would be energetically dominant even if numerically negligible in the wind.

One of the most important physics issues connected with the origin of the radio emitting particles is the fact that this origin is directly connected to the value of the pair multiplicity $\kappa$.
Estimates of $\kappa$ change by approximately two orders of magnitude depending on whether radio emitting particles  are considered as part of the pulsar outflow or not. 
The first MHD models and high-energy observations of the Crab suggested that $\kappa\sim 10^4$, assuming that only particles responsible for high-energy emission would be part of the pulsar outflow, while radio emitting particles are considered to be of different origin, possibly generated in a primordial outburst of the pulsar or elsewhere in the nebula \citep{Kennel:1984a, Atoyan:1996,Gaensler:2002}. It is also important to notice that in this scenario the hadronic component of the wind, if present, is expected to be energetically dominant.
Interestingly in the case of Crab these would be PeV ions.

On the contrary, if radio particles are considered as part of the pulsar outflow, the expected pulsar pair multiplicity is much higher, namely $\kappa\sim 10^6$, in which case the energy flux carried by a possible hadronic component would be completely irrelevant \citep{Bucciantini:2011}.

It is thus clear that discriminating between the two scenarios is of fundamental importance for a correct modelling of the pulsar wind composition.

\subsection{Particle acceleration mechanism at the wind TS}
It is a matter of fact that the Crab nebula is an efficient particle accelerator, with evidence of particles accelerated up to PeV energies \citep{Arons:2012}. The puzzling issue with this is the fact that relativistic magnetized shocks, as the TS is, are very hostile environments for particle acceleration, and the mechanism at the basis of the observed particle spectra is still a mystery.

The observed spectrum of the Crab nebula, similar to other PWNe, seems to suggest different acceleration mechanisms at work for low- and high-energy emitting particles, since it can be roughly modelled by assuming a broken power-law distribution function of the injected ultra-relativistic emitting particles, $N(E)\propto E^{-p}$, with $p\sim 1.5$ at low energies (for particles with $E\lesssim 100$ GeV) and $p\sim 2.2$  at higher ones (up to a few TeV).

The very flat slope of the radio emitting component is compatible with driven magnetic reconnection at the TS. 
This mechanism was investigated by \citet{Sironi:2011}, who performed particle-in-cell (PIC) simulations of a striped relativistic wind impinging on a shock, a structure similar to what is expected to occur at the pulsar wind TS. They assume that the striped morphology of the pulsar wind is maintained all the way to the TS, where compression leads the field to reconnect. Many reconnection islands form in the flow, where the unscreened electric field is able to accelerate particles. The properties of the resulting spectrum depend on the flow magnetization $\sigma$ and on the ratio between the wavelength of the stripes and the particles Larmor radii. 
This ratio is connected to the value of the pulsar pair multiplicity $\kappa$, and the result in the case of the Crab nebula is that in order to reproduce the observed radio spectrum, a magnetization $\sigma\gtrsim 30$ and a multiplicity of $\kappa \gtrsim 10^8$ are needed.
Such large multiplicity is well above the current predictions of pulsar magnetosphere models \citep{Timokhin:2015}, and in any case it makes it difficult to draw a self-consistent picture of the system, since a wind with the required high number of pairs is expected to undergo efficient dissipation well before the TS \citep{Kirk:2003}, destroying the stripes before the forced reconnection might start to work \citep{Amato:2014}. 

One possibility to alleviate the requirements on $\kappa$ is to accelerate particles near the polar cusps of the TS, where the shock front is closer to the pulsar, and as a consequence, the particle density is much higher (it decreases as $\sim 1/r^2$ with distance from the pulsar). However, on the other hand, no stripes are expected to be present at these locations unless the pulsar is an almost orthogonal rotator, {\it i.e.} with the angle between the magnetic and rotation axes close to $90^\circ$.


A different mechanism must be invoked to account for the much steeper high-energy component of the spectrum. A particle injection with slope $p \simeq 2.2$ is in fact what one expects from diffusive shock acceleration (DSA), i.e. Fermi~I-like acceleration at relativistic shocks. 
Contrary to driven magnetic reconnection, DSA is effective only when the magnetization is sufficiently low, in particular $\sigma \lesssim 10^{-3}$ \citep{Sironi:2009}. 
Since MHD simulations of the Crab nebula lead to values of $\sigma$ of at least one order of magnitude greater, DSA can only be effective in a few sectors of the shock, where magnetic dissipation is sufficiently strong to ensure that locally the plasma is almost unmagnetized.
It is not clear however if a sufficiently large fraction of the flow satisfies this condition. This again depends on the width of the striped region and on the efficiency of magnetic dissipation. According to current results of the 2-D MHD simulations that best reproduce the Crab nebula X-ray morphology, only few \% of the wind energy would be carried by a sufficiently low $\sigma$ flow \citep{Amato:2014}.

An alternative acceleration mechanism that received some attention in the literature \citep{Hoshino:1992, Amato:2006} is resonant absorption of ion-cyclotron waves in an ion-doped plasma (RCA hereafter). Its viability poses no special requirements on the wind magnetization, but does require that most of the energy of the wind is carried by ions.
The idea is that pairs would be accelerated by resonant absorption of the cyclotron radiation emitted by ions in the wind, when these are set into gyration by the enhanced magnetic field at the shock crossing. The outcome of this acceleration process in terms of efficiency, maximum achievable energy and accelerated particle spectrum depends on the relative amount of wind energy that is carried by ions. In order for RCA to provide any acceleration, the ions must be energetically dominant and therefore the mechanism is not viable if the multiplicity is larger than the pair to ion mass ratio, as would typically be the case if radio emitting electrons are part of the pulsar outflow \citep{Amato:2014}. On the other hand, it is not straightforward to make a connection with the flow properties that are relevant for MHD modelling and single locations, along the shock front, where RCA is more likely to work. This is why in the following we will mostly consider the first two processes discussed above when addressing the question of what MHD simulations can tell us about the acceleration sites of particles in different energy ranges.

\subsection{The gamma-ray flares}
\label{intro-gamma}

For a very long time the Crab nebula was believed to be a very stable source at both X-ray and gamma-ray energies, and hence it was considered as \emph{the} standard candle and calibrator for high-energy astrophysics  instrumentation. 
This was one of the reasons why the evidence of its gamma-ray flaring activity, first observed in September 2010 by the AGILE satellite, and then confirmed by FERMI, was so astonishing to the community \citep{Tavani:2011, Abdo:2011}.
Additional similar flaring events with a sudden release of an isotropic energy of $E_\gamma \sim 10^{41}$~erg have then been observed in the following years, confirming that the Crab is not a stable source at all \citep{Striani:2011,Striani:2013,Buehler:2012,Buehler:2014}.

During the flaring phase, the gamma-ray emission is enhanced dramatically in the spectral range between 50-100~MeV and a few GeV, basically the region where the steady-state synchrotron radiation drops and the IC contribution rises. 
The newly discovered non-thermal emission seems best modelled as due to a quasi-mono-energetic particle distribution, or a relativistic Maxwellian with Lorentz factor of order a few $10^9$ \citep{Buehler:2012}. We are thus in the presence of an extremely powerful cosmic accelerator, reaching energies above 1 PeV!

The timing of these flares is quite impressive. The increase in the gamma-ray light curve can be as fast as just a few hours, while the whole flares typically last from a few days up to a couple of weeks. Together with the extreme energetics involved (luminosities beyond 100~MeV up to $L_\gamma\sim 10^{36}$~erg s$^{-1}$), this clearly indicates that the source of the emission must be located in the innermost region of the Crab nebula, most probably around the TS. Unfortunately, in spite of several attempts, no clear indication of a parent candidate location has been found yet, among the variable features observed in the optical and X-ray bands (e.g. the knots): none of these shows any special variation, during the gamma-ray flares \citep{Weisskopf:2013,Rudy:2015, Bietenholz:2015}.

The current favorite physical interpretation of the gamma-ray flares currently relies on acceleration of particles in an electric field beyond the ideal limit, since the constraint $E \le B$ of relativistic MHD leads to an emission cutoff at approximately $200$~MeV. The obvious choice among the non-ideal effects is certainly that of sudden and powerful reconnection events in a magnetically dominated plasma, as appropriate for some regions in the post-shook nebular flow, in current sheets with sizes corresponding to the variable features observed in optical and X-ray bands in the innermost regions of the Crab nebula.

As far as particle acceleration is concerned, state-of-the-art 2-D and 3-D particle-in-cell numerical simulations of reconnection in relativistic pair plasmas, including radiative feedback, confirmed the possibility of acceleration of particles, with the correct energetics, higher magnetic fields ($\sim 1$~mG) and slope, in the narrow reconnection layers (current sheets) \citep{Cerutti:2014}.
Acceleration is allowed beyond the radiation reaction limit, with a predicted flux in the gamma-rays above 100~MeV exceeding the quiescence level by factors 2-3, as required to reproduce the flare events in the Crab nebula \citep{Cerutti:2013,Cerutti:2014}. The dynamics is dominated by driven tearing and by drift-kink instabilities, and the flare event is seen when the beam of accelerated particles is close to the assumed line-of-sight in the periodic simulation box \citep{Sironi:2014,Kagan:2015,Lyutikov:2016,Sironi:2016,Yuan:2016}.

\section{Results from 2-D simulations}
The observed complex X-ray morphology of PWNe, characterized by the presence of an equatorial torus and polar jets, was accounted for by different groups by means of 2-D MHD axisymmetric models, as a natural consequence of the pulsar wind anisotropy and striped structure \citep{Del-Zanna:2004,Komissarov:2004,Bogovalov:2005}.
Moreover the high-energy emission features were reproduced with great accuracy down to very fine detail \citep{Del-Zanna:2006, Camus:2009}, as well as polarization \citep{Bucciantini:2005a} and the complete emission spectrum \citep{Volpi:2008}. 
In this latter work the spectrum was computed including the gamma-ray emission component, due to the IC scattering process. 
The IC spectrum was computed by considering the extended Klein-Nishina differential cross-section and various photon targets for relativistic particles, namely synchrotron emitted photons, thermal photons coming from dust emission and photons from the cosmic microwave background. 
Particles responsible for low-energy emission (radio particles) and high-energy emission (optical/X-ray particles) were considered as part of two distinct families, the first extending up to an energy of approximately 100 GeV and the second starting from an energy approximately 2-3 times larger. This particle spectrum automatically ensures the appearance of a spectral break in the emission spectrum between the infrared and the optical bands. 
A second break, at ultra-violet (UV) frequencies, also arose naturally from the synchrotron burn-off effect.

It was pointed out in that work that the reduced dimensionality of the simulation reflects in an artificial compression of the magnetic field around the polar axis, such that in order to reproduce the high-energy morphology of the Crab nebula a smaller value of the average magnetic filed is found ($\sim50\, \mu$G), which results to be approximately $3\div 4$ times smaller than that estimated from the integrated emission spectrum \citep{de-Jager:1992}. With this low average field, electrons are affected by weaker synchrotron losses and a steeper spectral index (of approximately $0.8$) is required in order to fit the high-energy component of the spectrum. At the same time, with a lower field, a larger number of particles is needed in order to reproduce the synchrotron emission and this reflects into an overestimate of the IC component by the corresponding factor. 
The conclusion of that work was that, in order to reproduce the whole spectrum self-consistently, a larger and more diffuse magnetic field is mandatory, and 3-D models are indeed needed.

Two-dimensional MHD models have also dealt with the question of time variability at different wavelengths, showing that wisps can be reproduced in terms of time scales, morphological patterns and outward velocities \citep{Camus:2009,Olmi:2014}. 
In these models wisps originate very close to the TS shock front, in the region where eddies of high velocity develop.

\subsection{Investigating the origin of radio emitting particles}
For many years, MHD numerical models have mainly been concerned with the properties of high-energy emission, while ignoring emission at lower energies.  As previously mentioned, clarifying the origin of radio emitting particles is fundamental to obtain a correct estimate of the value of the pulsar pair multiplicity $\kappa$ and, as a direct consequence, constraints on the particle acceleration process.

In a recent study \citet{Olmi:2014} have considered different possibilities for the origin of radio particles, testing whether it is possible or not to discriminate their origin based on the resulting emission morphology. 
The question is whether radio particles are part of the pulsar outflow, and accelerated at the TS together with high-energy emitting particles, or if they are of different origin and accelerated somewhere else in the nebula.

Three different hypotheses were then considered: radio particles are injected in the nebula as part of the pulsar outflow (case A); radio particles are uniformly distributed in the nebula, produced somewhere else, for instance in the thermal filaments, and continuously accelerated by the interaction with local turbulence (case B); radio particles are a relic population, born in a primordial outburst of the pulsar during its early life, and still present thanks to their very long lifetimes against synchrotron losses (case C).

\begin{figure}
\centering
 	\includegraphics[scale=0.65]{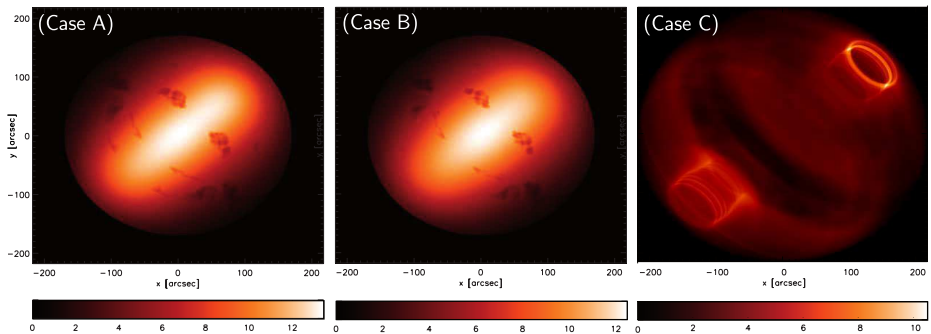}
\caption{ Radio intensity maps at $\nu=1.4 $ GHz, with intensity given by the color bar in units of mJy/arcsec$^2$. Emission is computed at $\sim1000$ yr and is plotted in linear scale and normalized to its maximum value. \emph{Left panel}: radio emitting particles are injected as part of the pulsar outflow into the nebula (case A). \emph{Central panel}: radio particles are considered as uniformly distributed in the nebula (case B). In all cases the maps are treated as in \citet{Bandiera:2002}, subtracting small scale structures.
\emph{Right panel}: radio particles are injected in the nebula at early times and advected with the flow, without diffusion or any further acceleration (hypothesis C).} 
\label{fig:radioTOT}
\end{figure}

Each scenario was tested by simulating the entire evolution of the nebula, with the suitable numerical tracers following the energy evolution of radio particles. Emission properties were then computed on top of the numerical data corresponding to an age of the Crab nebula of approximately 1000 yrs.
Results obtained in the three different scenarios are shown in Fig.~\ref{fig:radioTOT}.
When particles are injected only for a short time during the initial stages of the nebular evolution (case C), with the injection limited to $t\le 100$ yr, emission maps are clearly different from what observations show, as can be seen from the right-most panel of Fig.~\ref{fig:radioTOT}.
Here radio emission mainly comes from the outer shell of the nebula, while the inner region results to be much fainter than observed. 
This is the result of the advection of radio particles during the early stages of evolution. 
Particles are advected by large-scale flow vortexes to the outer region of the nebula, and the absence of freshly injected particles at later times results into a lack of emitting particles in the central region, since all of them are confined to the outer shell.
Kinetic diffusion, which is not included in the model, is however expected to have negligible effects on these low-energy particles, while turbulent flow diffusion might be more relevant.
The inner nebula can be somehow enriched in primordial electrons by the effect of the Rayleigh-Taylor (RT) instabilities that set in at its outer boundary \citep{Porth:2014a}. In the discussed simulations the resolution is not sufficiently high to allow this mechanism to produce significant mixing, and effectively enrich the inner nebula with primordial electrons. In any case we do not expect the effects of turbulent RT mixing to be sufficient to reconcile our case C with radio observations (see e.g. \citet{Bandiera:2002}), which require that the electrons fully concentrated in the outer layers should become spread more or less uniformly in the whole nebula, down to the very inner regions, where the filaments do not even reach (they are seen to penetrate inside to 1/3 of the radius).

\begin{figure}
\centering
 	\includegraphics[scale=0.7]{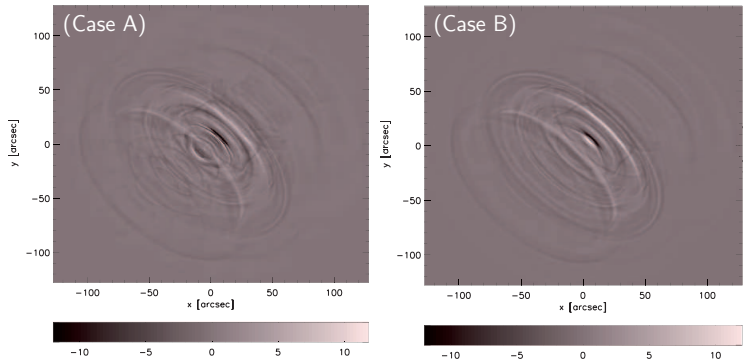}
\caption{Appearance of wisps at radio frequency in case A (left panel) and case B (right panel). Wisps maps are obtained by subtracting images of the emission separated by a time interval of about 2 months. Outward motion of  the wisps can be observed as consecutive darker and lighter arcs.} 
\label{fig:origin2}
\end{figure}

Therefore the hypothesis of an early burst is not a viable description of the origin of the radio emission from the Crab nebula. 
A caveat to this conclusion is that it is obtained by means of 2-D simulations. While in the inner region of the nebula we expect the 2-D description to be a good approximation of the dynamics, given that the kink instabilities that are suppressed by the reduced dimensionality are unlikely to have the time to develop, on the larger scales the dynamics is shown to be rather different in three dimensions, as already mentioned. 
Notice that the axisymmetric model is robust within a few TS radii, while moving further from the symmetry axis, the differences with the full 3-D model become more and more remarkable. A more detailed discussion on this can be found later on in section \ref{sec:3Dm}.
The possibility exists that in 3-D case C will look similar to our current case B, thanks to more effective mixing and wandering of magnetic field lines. It is clear that an analogous study will need to be done in three dimensions before the final word against a primordial origin of the radio emitting particles can be spelled.
On the other hand cases A and B lead to almost undistinguishable emission maps, and comparison with observations does not allow one to discriminate between the two cases.
Moreover, the analysis of the wisp time variability in the inner region of the nebula is also found to be very similar in the two cases (Fig.~\ref{fig:origin2}).
Wisps appear with similar features and outward velocities whether radio particles are considered as uniformly distributed in the nebula or injected and accelerated at the TS as the high-energy ones. 

This proves that the existence of wisps does not imply that emitting particles are injected and accelerated at the shock. In fact wisps arise in the simulated maps as a consequence of the intrinsic properties of the MHD flow: bright moving features are associated with local enhancements of the magnetic field and Doppler boosting of material moving toward the observer.

\subsection{Investigating the particle acceleration mechanism at the shock}\label{subs:acceleration}
%
Within a MHD description wisps arise thus as a natural consequence of the underlying flow properties.
Therefore the fact that wisps are observed to be not coincident at the different wavelengths suggests a difference in the acceleration sites of the particles responsible of such emission. 

As previously mentioned, different acceleration mechanisms require very different physical conditions to be effective, and thus identifying the location at which particles have been accelerated can put strong constraints on the mechanism at work.

In \citet{Olmi:2015} different possibilities were considered for acceleration of radio, optical and X-ray particles. 
In particular two distribution functions were defined for accounting for radio and X-ray particles. Since the distribution functions were assumed in such a way as to fit the complete spectrum of the Crab nebula (i.e. cutoff energies and normalization constants are chosen in order to obtain the best fit of the spectrum), the optical component is automatically determined as a superposition of the radio and X-ray contributions.

Three different hypotheses were considered for injection of particles: in case \textit{I} particles are injected uniformly at the shock surface; in case \textit{II} particles are injected in a wide equatorial sector or in the complementary narrow polar one; in case \textit{III}, emitting particles are injected in a narrow equatorial region, or in the complementary wide polar one.

The entire evolution of the Crab nebula was then simulated by introducing as many numerical tracers as necessary in order to follow the evolution of particles injected at the different locations. 
In particular, each family could be injected in five different zones: wide or narrow polar caps, wide or narrow equatorial belt, or along the entire TS surface. 
The entire system was evolved up to the actual age of the Crab ($\sim 1000$ years), and wisps' profiles were extracted from synthetic intensity maps during a period of 10 years at the end of the evolution, with monthly frequency. 

For a direct comparison with wisps' observations published by \citet{Hester:2002}, \citet{Bietenholz:2004} and \citet{Schweizer:2013}, emission maps were convolved with the appropriate instrumental point spread functions (PSFs). The adopted full width at half maximum (FWHM) was: $0.5\arcsec$ for X-rays (from Chandra), $1.8\arcsec$ for radio (Very Large Array) and $0.75\arcsec$ for optical emission (North Optical Telescope).

Following the analysis by \citet{Schweizer:2013} intensity peaks were extracted from a $3\arcsec$ wide slice around the polar axis in the upper hemisphere of each map, where wisps are more prominent. For ease of comparison only peaks with intensity greater than $1/3$ of the intensity maximum were taken into account. Results are shown as plots of the radial distance (from the pulsar, in arc seconds) of the local intensity maxima as a function of time (in months).

\begin{figure}
\centering
 	\includegraphics[scale=0.64]{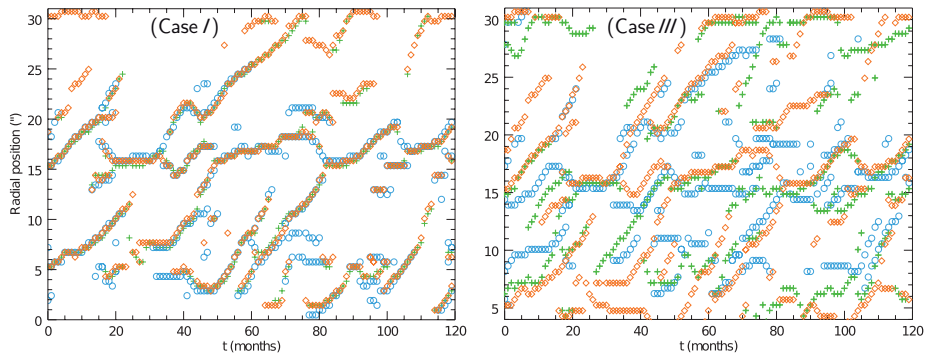}
\caption{Radial positions of the local intensity maxima (in arcseconds) as a function of time (in months) with orange diamonds identifying radio wisps ($\nu_r=5$ GHz), green crosses optical ones ($\nu_o=3.75 \times 10^{14}$ Hz) and light-blue circles for X-rays (1 keV). 
On the left case \textit{I} is shown. On the right case \textit{III} is shown, with X-ray particles injected in the equatorial zone and radio ones injected in the complementary sector.} 
\label{fig:wisp1}
\end{figure}

In Fig.~\ref{fig:wisp1} wisps obtained in the case of uniform injection or injection in a wide (narrow) equatorial (polar) region are compared. As expected, when particles of different families are injected at the same location, i.e. across the entire shock surface as in case \textit{I}, wisps appear to be coincident at the different wavelengths. But as soon as they are injected at distinct locations (case \emph{II} and \emph{III}) wisps at radio, optical and X-ray frequencies are no more coincident. 
Here we only show wisps obtained under case \emph{III} hypothesis (with X-ray particles injected in the equatorial region and radio ones in the opposite polar zone), while for a complete discussion of the wisp properties obtained in the different scenarios we refer to \citet{Olmi:2015}. 
The main result of the analysis presented in the cited paper is that, in order to reproduce the absence of X-ray wisps in the region within $\sim 6\arcsec$ from the pulsar, as observed in \citet{Schweizer:2013}, X-ray particles must be injected in a narrow equatorial sector, approximately coincident with the striped region of the wind where dissipation of magnetic field is expected to be most efficient.

This may indicate that X-ray particles are produced via Fermi I acceleration in the striped zone, where magnetization can be lowered enough by dissipation so as to allow this mechanism to be effective.

In \citet{Schweizer:2013} it was also found that optical wisps are characterized by narrower angular profiles than X-ray wisps. Already in that article, two possible explanations were suggested for this finding: either a stronger Doppler boosting of the optical emitting region, or a higher degree of anisotropy of the optical emitting particles. 
A relevant investigation in this sense was carried out by \citet{Yuan:2015}, who showed that very different emission patterns can result from different assumptions on the emitting particle pitch angle distributions.
In the MHD modelling by \citet{Olmi:2015} the particle distribution function was assumed to be isotropic at all energies and the model was found to correctly reproduce the angular size of the X-ray wisps, but overestimated the extension of the optical ones. This favours the idea that if the wisps are to be explained as a result of the properties of the MHD flow alone, then the X-ray emitting particles must be more isotropic than optical emitting ones, a fact that can be used in principle to put constraints on the acceleration mechanism, though this goes beyond the scopes of the present work.

Moving to lower frequencies, strong constraints on radio emission are again difficult to derive: the only case that can be easily excluded is the one in which radio particles are injected in a narrow polar cone, since no wisps appear to be produced.
Since radio wisps must not be coincident with X-ray ones, the remaining possibilities are that acceleration of low-energy particles happens in the complementary polar region or in a wider equatorial stripe.

A possible interpretation of this finding is that radio particles could result from driven magnetic reconnection primarily occurring at moderately high latitudes, where conditions for driven magnetic reconnection to operate as an acceleration mechanism might be locally satisfied.

Going beyond the qualitative conclusions this study has provided requires an effort to compute the particle spectrum that the study by \cite{Sironi:2011} would predict as a function of latitude along the shock surface for a given value of the inclination between magnetic and rotation axis. In the meantime, however, recent developments in pulsar and pulsar wind theory \citep{Philippov:2015, Tchekhovskoy:2016} are somewhat modifying the picture of the wind that was assumed within the modelling summarized above, showing that the energy flux from the pulsar has a stronger latitude dependence than previously thought. 
Early results from 3-D modelling have shown that also our assumptions on the level of magnetization at the shock need to be modified, allowing for much more magnetized winds than 2-D models (see the results and the discussion in the next section). 
All this wealth of information will have to be incorporated in the models to obtain a reliable assessment of the particle acceleration process/es at work in the different energy ranges.
\section{A step forward: 3-D models}\label{sec:3Dm}
%
The 2-D modelling of PWNe, while very successful at reproducing the morphology of the inner region of these nebulae, is clearly not able to provide at the same time a correct description of the nebular magnetic field on the large scales, as demonstrated by the lack of agreement between the simulated emission spectrum and multiwavelength observations.
In 2-D models, the lack of the third degree of freedom suppresses the kink mode, and the magnetic field is allowed to artificially pile up around the symmetry axis. As a consequence, the value of the wind magnetization parameter cannot be raised above $\sigma\simeq 3\times10^{-2}$, otherwise the effect of the compression affects also the evolution of the TS, reducing its size to much smaller values than observed. The small magnetization of the wind that one is forced to assume also leads to an average value of the field in the nebular volume which is lower than  that inferred from spectral modelling ($\sim200 \,\mu$G) by a factor of $3\div4$.
As already discussed, this strongly affects the properties of the nebula at large scales, such as the total integrated spectrum and the emission maps of the entire nebula.
The first 3-D models of the Crab nebula via relativistic MHD simulations were presented by \citet{Porth:2013, Porth:2014}. 
Since these simulations are computationally very demanding, they could only evolve the system for $\sim 70$ years, so that the self-similar expansion phase was not completely reached.
Nevertheless, many interesting results were found. Working in three dimensions allowed the authors to increase the magnetization to values greater than unity, since now the kink instability is free to set in and leads to efficient  dissipation of the field inside the nebula, as already noticed in previous simulations of jets \citep{Mignone:2013}. 
Notice that in these simulations the dissipation is only numerical.
They also found that the innermost structure is perfectly analogous to that obtained with 2-D axisymmetric models (of lower magnetization), indicating that MHD models are robust in describing the jet-torus morphology of PWNe. Jets are retrieved in the downstream region of the TS as the result of flow collimation by magnetic hoop stresses. The loss of axisymmetry reflects in a complex structure of the field in the simulated PWN: the highly ordered field no longer survives in the nebula, but it becomes more or less randomized. The toroidal component of the field is still dominant near the TS, while around the polar jets and close to the PWN boundaries (the contact discontinuity with the supernova ejected material) the poloidal component becomes more important.

Despite the important results introduced by this work, some problems are still unsolved.
In particular, the value of the magnetic field extrapolated to the actual age of the system appears to be well below what deduced observationally (Porth, \emph{private communication}). This unfortunately leaves open the question of whether the comparison of emission maps and spectra with real data will be satisfactory.
However, it is clear that these kind of 3-D simulations are the primary tool to reach a comprehensive description of PWNe in all their aspects.

%
We have recently performed our own 3-D simulation of the Crab nebula, with the aim of reaching at least the self-similar expansion phase, if not to follow the complete evolution of the system ($\sim 1000$ years). Since the simulation by \citet{Porth:2014} shows a faint emission in the torus region, and a complete absence of emission from the polar jets, we have decided to use a slightly different model for the injected pulsar wind, with the main goal of reproducing the observed emission from the Crab nebula, as we discuss below. For ease of comparison with Porth's results, we maintain some initialization choices, namely at $t=0$ wind conditions are defined for $r \le r_\mathrm{wind}=1$~ly, the slowly expanding supernova ejected material extends up to $r_\mathrm{ej}=5$~ly, whereas static interstellar medium (ISM) conditions apply at larger radii, up to the simulation box boundaries. This setup corresponds to an initial age of the system of $t_0\simeq250$ years, thus, in order to follow the complete evolution of the Crab nebula, approximately 700 years of simulated evolution would be needed. 
In the following we will refer to two different times: $t$ as the simulated time and $t_\mathrm{eq}= t + t_0$ as the equivalent evolution time of the system.

Our simulation is performed with the shock-capturing, finite-volume numerical code Pluto \citep{Mignone:2007a, Mignone:2012}, which supports adaptive mesh refinement (AMR), and it has been run on 2048 processors at the Tier0 CINECA (Bologna, Italy) FERMI cluster.  
We define a cubic numerical grid of $\left[-10\,\mathrm{ly}, +10 \,\mathrm{ly}\right]^3$, that can contain the whole nebula up to the end of its evolution, with the pulsar wind continuously injected in a sphere with radius $0.01$~ly (which allows the TS to detach from the inner boundary of the wind). 
Seven AMR levels are employed to guarantee the required accuracy close to the centre of the domain, where the wind is injected. 
The resolution is then decreased dynamically with distance from the pulsar, preserving a higher resolution around the wind TS and in the inner part of the nebula, more or less in the region in which the jet-torus morphology develops. Since the TS is not static but its position changes with time, we cannot use a static decomposition of the domain, and AMR is a necessary requirement.

The solenoidal condition of the magnetic field ($\vec{\nabla}\cdot \vec{B}=0$) is ensured by the \textit{divergence-cleaning} method, which augments the relativistic MHD system with an extra equation for a generalized Lagrange multiplier, allowing the propagation and progressive damping of the divergence errors \citep{Dedner:2002}. 
The high Lorentz factor flow is managed by the selective use of the Harten-Lax-van Leer (HLL) Riemann solver in the most relativistic regions, or the HLLD (where \enquote{D} stands for discontinuities) solver \citep{Mignone:2009} for the smoother part of the system.  Additional smoothing in the proximity of the strong shock is achieved by a shock flattening option.


\begin{figure}
\centering
 	\includegraphics[scale=0.38]{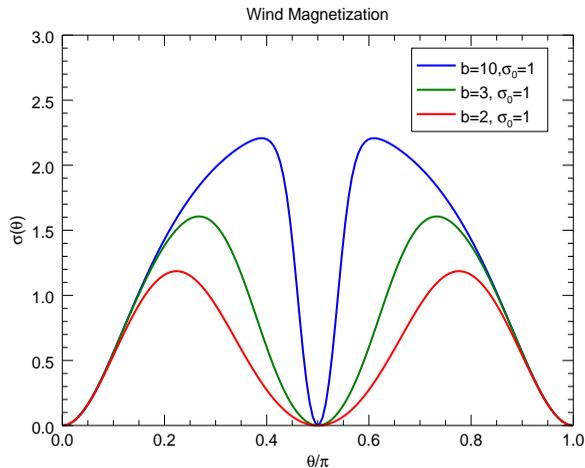}
\caption{
Shape of the wind magnetization $\sigma(\theta)$ for the value of $\sigma_0=1$
employed in our 3D simulation and for three values of b. The upper (blue)
curve is the case with a narrow striped wind region used here, with $\sigma_\mathrm{eff}=1.5$,
while the bottom (red) curve refers to a case that approximately reproduces 
the magnetization function as in the B3D run in \citet{Porth:2014}.
} 
\label{fig:sigma_eff}
\end{figure}

As previously mentioned, the fundamental difference between our model and the one by \cite{Porth:2014} is in the wind parametrization and angular dependence.
Following our previous works in two dimensions \citep[see e.g.][]{Olmi:2014,Olmi:2015}, we assume an energy flux in the wind which is still axisymmetric and depends on radius $r$ and polar angle $\theta$ (measured from the $z$ symmetry axis, which is assumed to be the rotation axis of the central pulsar) approximately as predicted by the split-monopole model \citep{Michel:1973}. 
The observed jet-torus morphology and the expected oblate shape of the TS are accounted for by considering an anisotropic distribution of the energy flux in the wind. The level of anisotropy is governed by an appropriate free parameter of the model ( $\alpha$) which represents the ratio between the polar and equatorial energy fluxes in the wind, and which must be chosen $\alpha\ll 1$ \citep{Del-Zanna:2004}.

Since the numerical cost of 3-D simulations is very large, we did not include additional numerical tracers in the present simulations, with the exception of the tracer for the particle maximum energy (the $\epsilon_\infty$ defined in \citet{Del-Zanna:2006}), which is absolutely mandatory to account for the synchrotron emission. As a consequence we could only inject particles uniformly along the shock front.

The striped morphology of the magnetic field is ensured by choosing $B(r,\theta)\propto  f(b,\theta) \sin\theta$, where $f(b,\theta)$ is an appropriate function of latitude. $f$ also depends on the free parameter $b$, which represents the width of the striped zone of the wind. In particular, for large values of $b$, the pure split-monopole configuration is recovered, while in the case of $b\sim 1$, corresponding to a large striped region (of approximately $60\degr$), due to a highly tilted pulsar magnetosphere, dissipation before the TS is supposed to have shaped the field strength at basically all latitudes (see Fig.~\ref{fig:sigma_eff} and discussion in \citet{Del-Zanna:2006}). 
The initial magnitude of the magnetic field is governed by the initial magnetization $\sigma_0$, which is contained in the proportionality constant. 

The set of these three free parameters ($\alpha,\, b,\,\sigma_0)$ is chosen based on the model that appears to best fit the Crab observations in 2-D MHD simulations. In particular we use $b=10$ and $\alpha=0.1$, and, based on the findings of the previous 3-D study, we raise the magnetization to unity ($\sigma_0=1$).
The initial value of the wind Lorentz factor is chosen to be $\gamma_0=10$. While the actual wind Lorentz factor is certainly much larger, say up to $\sim 10^6$ \citep{Kennel:1984}, the adopted value is high enough to guarantee that the flow is highly relativistic, in which case the post-shock dynamics is known to be independent on the exact value of $\gamma_0$.

The chosen parametrization corresponds to an effective magnetization of $\sigma_\mathrm{eff}=1.5$, that can be obtained by integrating over $\theta$ the quantity $\sigma(\theta)$ defined in equation~(5) of \citet{Olmi:2014}. Different possible choices of $\sigma_\mathrm{eff}$ are shown in Fig.~\ref{fig:sigma_eff}, with our current one being blue in colour. 
This quantity can be directly compared with the quantity $\bar{\sigma}$ in Table~1 of \citet{Porth:2014}, noticing that our parametrization roughly corresponds to their case D3D, in which the pulsar obliquity is of order 10\degr.

\begin{figure}
\centering
 	\includegraphics[scale=0.68]{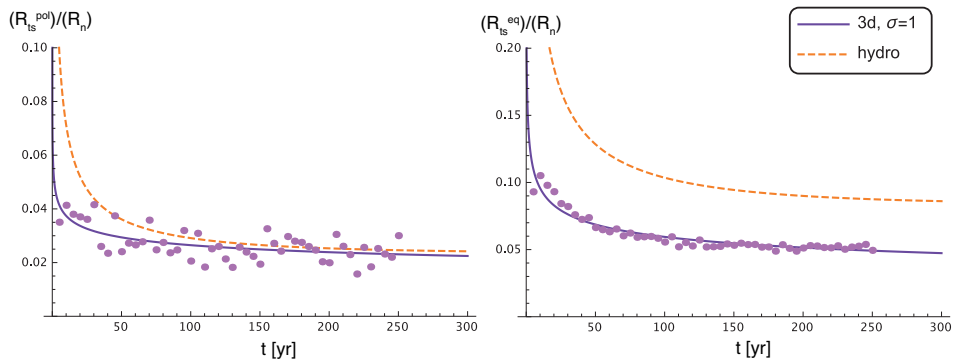}
\caption{Evolution with time of the ratio between the equatorial/polar radius of the TS and the radius of the nebula. Full circles represent data extracted from  the simulation, solid lines our fit to these data and dashed ones the expected behavior in the hydro-dynamical case, as a comparison. 
In the left (right) panel, the ratio between the polar (equatorial) radius of the TS and the outer radius of the nebula is shown. As can be noticed from these plots, the self-similar phase of evolution appears to be fully reached.} 
\label{fig:3d_RTSeq_rn}
\end{figure}

In Fig.~\ref{fig:3d_RTSeq_rn}, the time evolution of the nebula and its TS are shown at the simulation time of 250 yr, corresponding to the physical time $t_\mathrm{eq}\simeq 500$ yr.
In particular, in the left-hand side of the figure, the ratio between the polar radius of the TS, defined as the maximum extent of the TS in the polar direction, and the radius of the nebula is shown. 
Assuming a polytropic equation of state with index $4/3$ and a radial power law for the evolution of the nebula radius, the hydrodynamical evolution can be obtained from the momentum conservation at the TS (see equation~(44) from \citet{Porth:2014}). This trend is also drawn as a dashed line in the plots, as a basis for comparison. In the right-hand panel the same plot is produced for the ratio between the equatorial radius of the TS, defined as the maximum extent of the TS in the equatorial plane, and the radius of the nebula.
The fits to both ratios with a suitable function is also shown (as a solid line) to help the reader to see the long-term trend within the highly dynamical evolution.

As can be easily understood from these plots, the system seems to have fully reached the self-similar phase of expansion.

\begin{figure}
\centering
 	\includegraphics[scale=0.58]{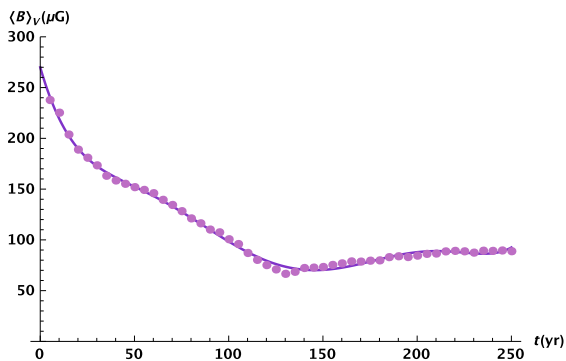}
\caption{Evolution with time of the magnetic field strength, averaged over the entire nebula volume. Filled circles represents data extracted from simulations, while the line is the fit to the data.
} 
\label{fig:3d_Bmean}
\end{figure}

In Fig.~\ref{fig:3d_Bmean} the evolution with time of the magnetic field strength averaged over the nebula volume is also shown. The magnetic field appears to decrease down to values of the order of $\sim 100 \, \mu$G in the first 100 years of simulation, which means that the average field in the nebula is approximately a factor of 2 lower than expected.
 Anyway the trend seems to indicate that the strong magnetic dissipation of the first period of evolution slows down after around 100 years: the field strength decreases down to a minimum value and afterwards, for the following 100 years, it stays almost constant, or it even increases somewhat. 
This might indicate that magnetic dissipation progressively becomes less important, leading the magnetic field to saturate to the equipartition value at some point of the evolution. 
For times up to those investigated by \citet{Porth:2014}, the trend we find for the magnetic field time evolution perfectly agrees with their results for the time evolution of the magnetic energy in the nebula.

\begin{figure}
\centering
 	\includegraphics[scale=0.26]{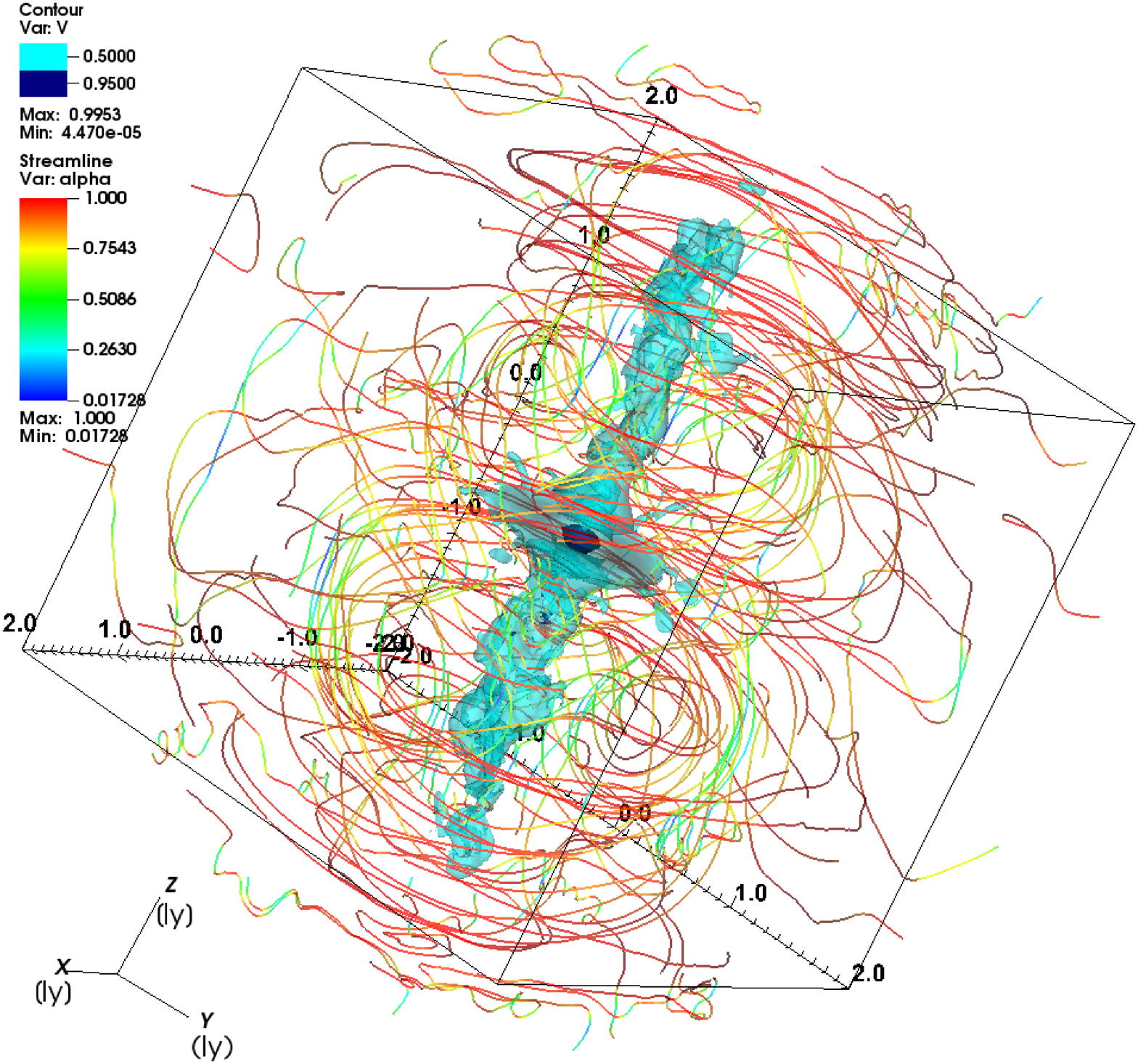}
\caption{Magnetic field lines drawn with seed points laying on a sphere of radius 2.5 ly, at the time t=250 yr. The contrast between toroidal and poloidal components ($\alpha_B=B_\mathrm{tor}/B$) is highlighted by the color scale, with red color meaning that the field is completely toroidal and the blue one that it is predominantly poloidal. 
The velocity field is also represented as a 3D contour plot, with color levels corresponding to $0.95c$ and $0.5c$. Dimensions of the nebula are indicated by the box axes, in units of ly.} 
\label{fig:3d_fieldL}
\end{figure}

As expected, after some time, the magnetic topology becomes more complex, with the development of a poloidal component. In Fig.~\ref{fig:3d_fieldL} the structure of magnetic field lines at t=250 yr is shown, with the colour bar indicating the ratio of the toroidal to the total field magnitude, measured by the quantity $\alpha_B=B_\mathrm{tor}/B$. Preponderance of different components of the field are expressed as red/yellow colours and green/blue ones, indicating, respectively, the toroidal and poloidal components. As expected, the field is mainly toroidal in the inner part of the nebula, as only this component is injected in the wind and amplified post-shock, while the poloidal component is mostly important outside along the polar jets and in the external regions, where the field structure is more efficiently modified by polar high-speed flow or by the turbulent motions, respectively. In the same figure, we also show isosurfaces of constant velocity, dark blue indicating the wind region and light blue to highlight the jets, where we have chosen a reference value of $v=c/2$.

\begin{figure}
\centering
 	\includegraphics[scale=0.18]{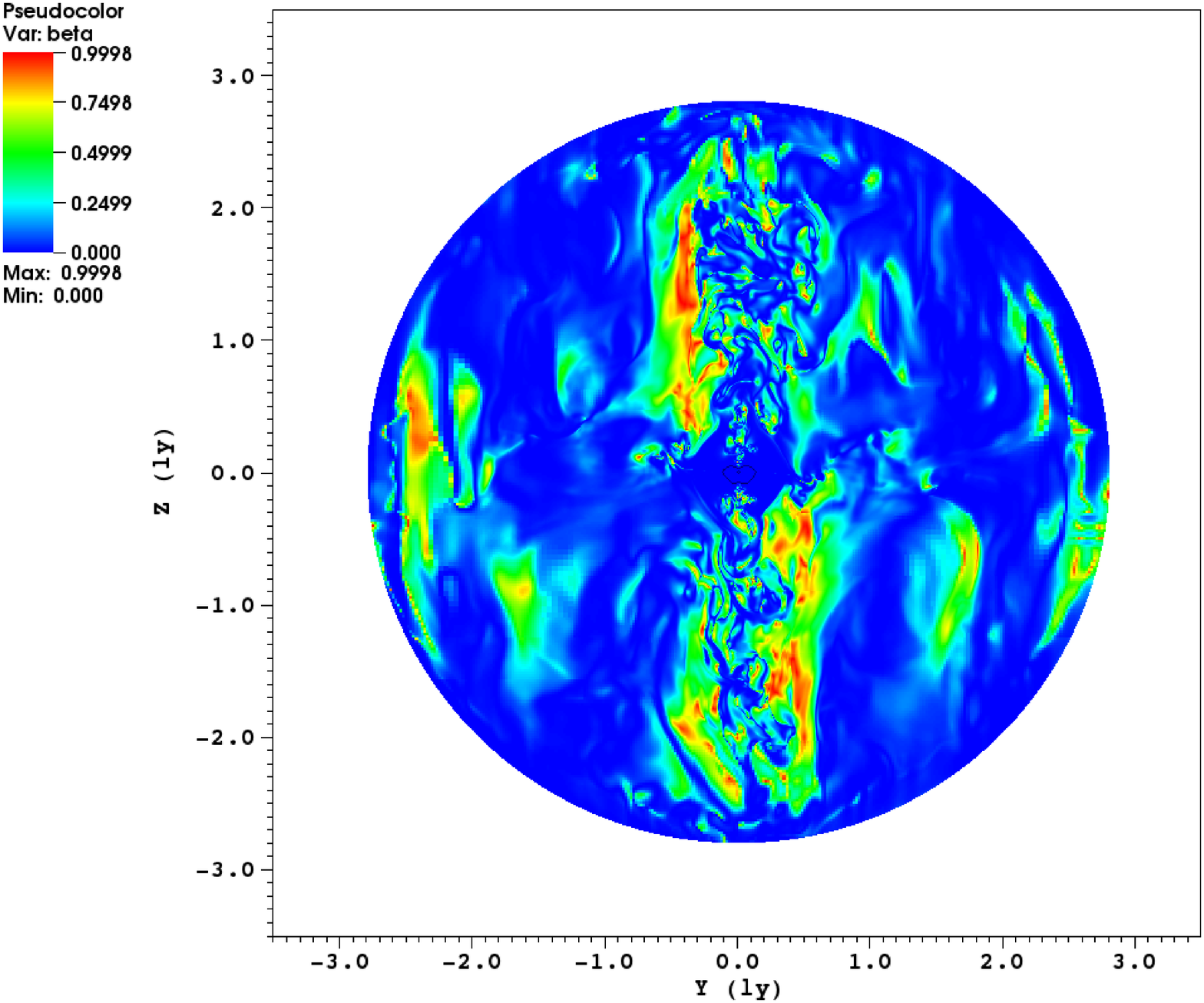}\,
	\hspace{-0.6cm}
	\includegraphics[scale=0.183]{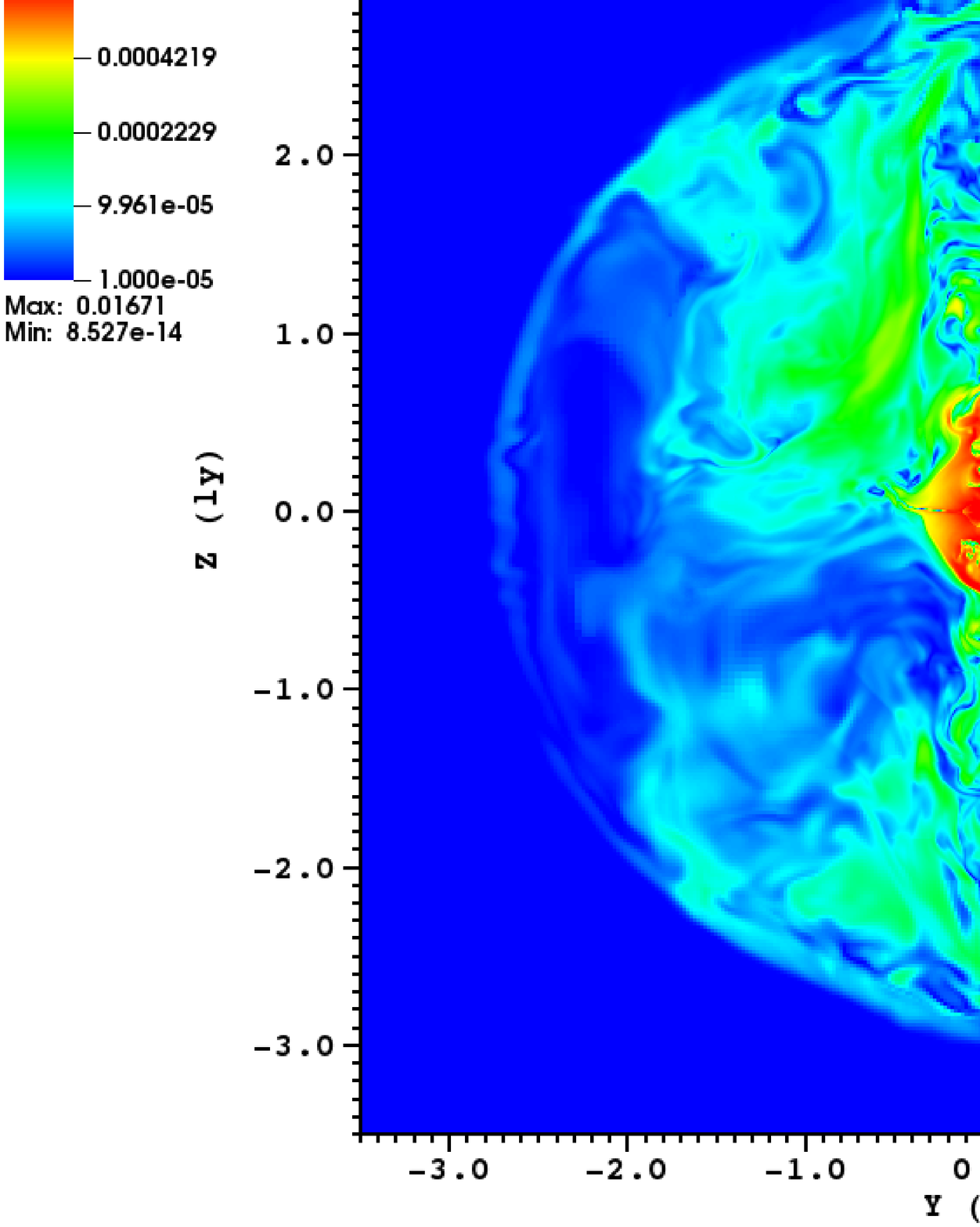}\,
\caption{\textit{Left panel}: The ratio of the polar to the total magnetic field magnitude in a 2D slice of the data cube at t=250 yr. Data are also cropped within a circle of radius equal to the nebular radius (2.8 ly) for ease of interpretation. As expected, the polar component is mostly important near the polar jets. \textit{Right panel}: image of the 2D distribution of the total magnetic field magnitude in the ${0,\, y,\, z}$ plane, shown in units of Gauss at the time 250 yr. The magnetic field appears to be stronger close to the TS, where the mean value is around 1 mG.} 
\label{fig:Bvarie}
\end{figure}

The relative importance of the poloidal and toroidal component of the field can also be seen in a 2-D slice of the data cube, shown in the left panel of Fig.~\ref{fig:Bvarie}. Here the variable $\beta_B \equiv B_\mathrm{pol}/B$ is represented at t=250 yr in the plane ${(0,\, y,\, z)}$, and red colour indicates complete dominance of the poloidal component of the magnetic field (i.e. $\beta_B\sim 1$).
As expected, the poloidal field is mostly concentrated near the polar jets. 
The variable $\beta_B$ is also a good marker for estimating the region where results obtained with 2-D axisymmetric models are robust.
In fact, where the ratio of the polar component to the total magnetic field is small (blue and green regions, where the polar component is below 50\% of the total) the axisymmetric description (where the field is completely toroidal) is reliable. In yellow and red regions the two descriptions are substantially different. This means that the 2-D description provides a good approximation within approximately $1/5$ of the nebula radius.
Looking at the panel on the right side of Fig.~\ref{fig:Bvarie}, it is evident that the magnetic field magnitude is stronger in the inner nebula, where the field is mostly toroidal, and where the average value is close to what is expected from spectral modelling $\left<B \right>\simeq 200 \, \mu\mbox{G}$.

\begin{figure}
\centering
	\includegraphics[scale=0.3]{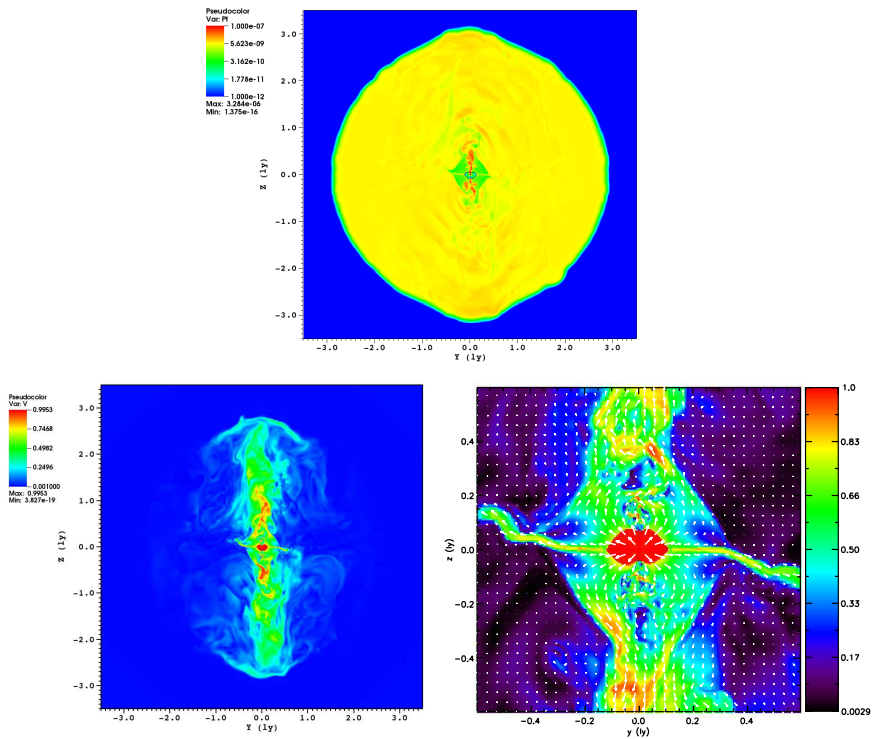}\,
\caption{\textit{Upper row}: 2D map of the particle pressure distribution, with intensity varying according to the color (logarithmic) scale. 
\textit{Left panel, bottom row}: Magnitude of the velocity field in a 2D slice at t=250 yr. 
\textit{Right panel, bottom row}: blow-up of the inner nebula in a plot of the velocity magnitude with arrows indicating the velocity field. Effects of the hoop stresses, guiding the flow in the polar direction to form jets, are clearly visible.} 
\label{fig:dynvarie}
\end{figure}

\begin{figure}
\centering
 	\includegraphics[scale=0.45]{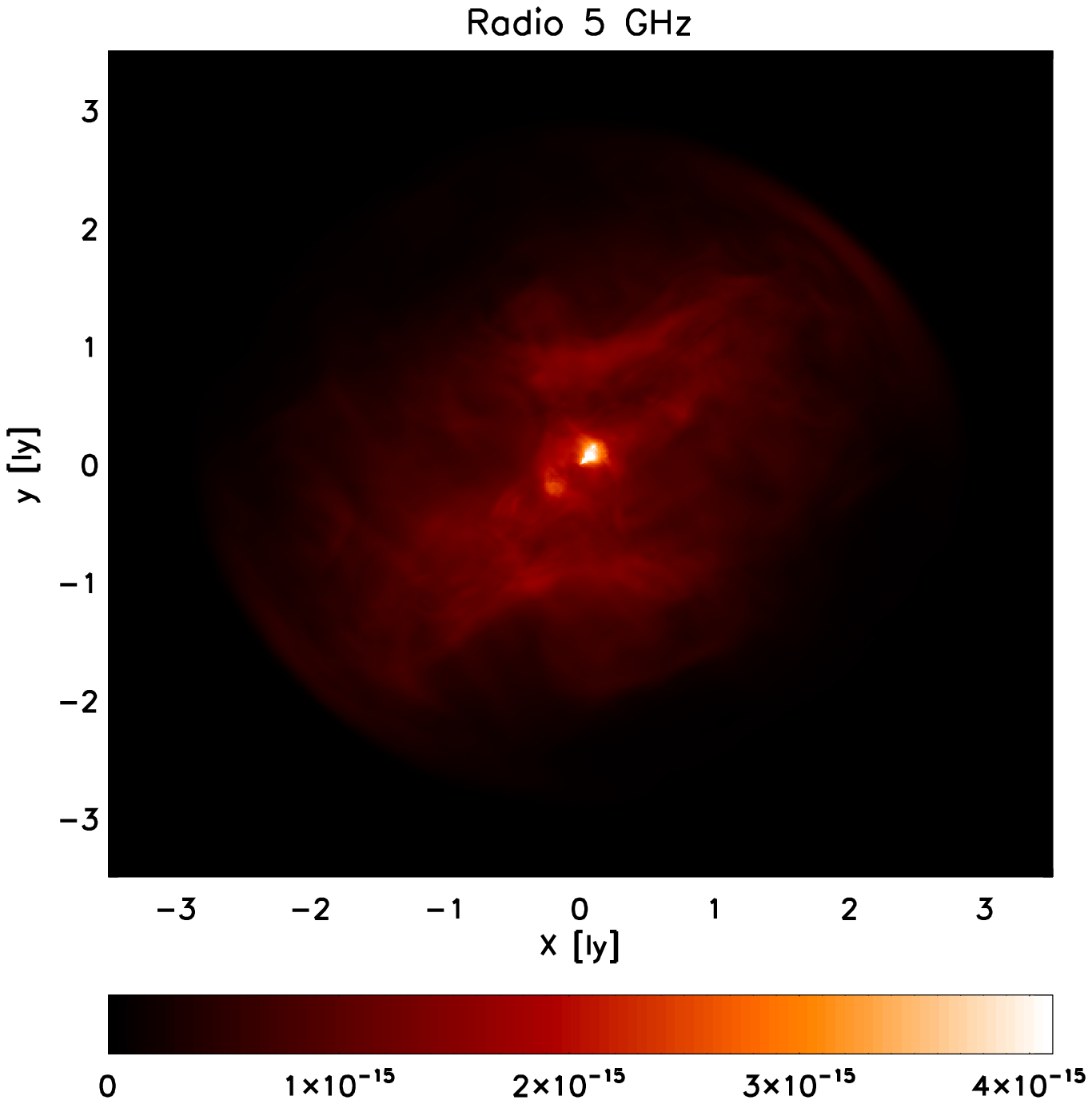}\,
	\hspace{-0.4cm}
	\includegraphics[scale=0.45]{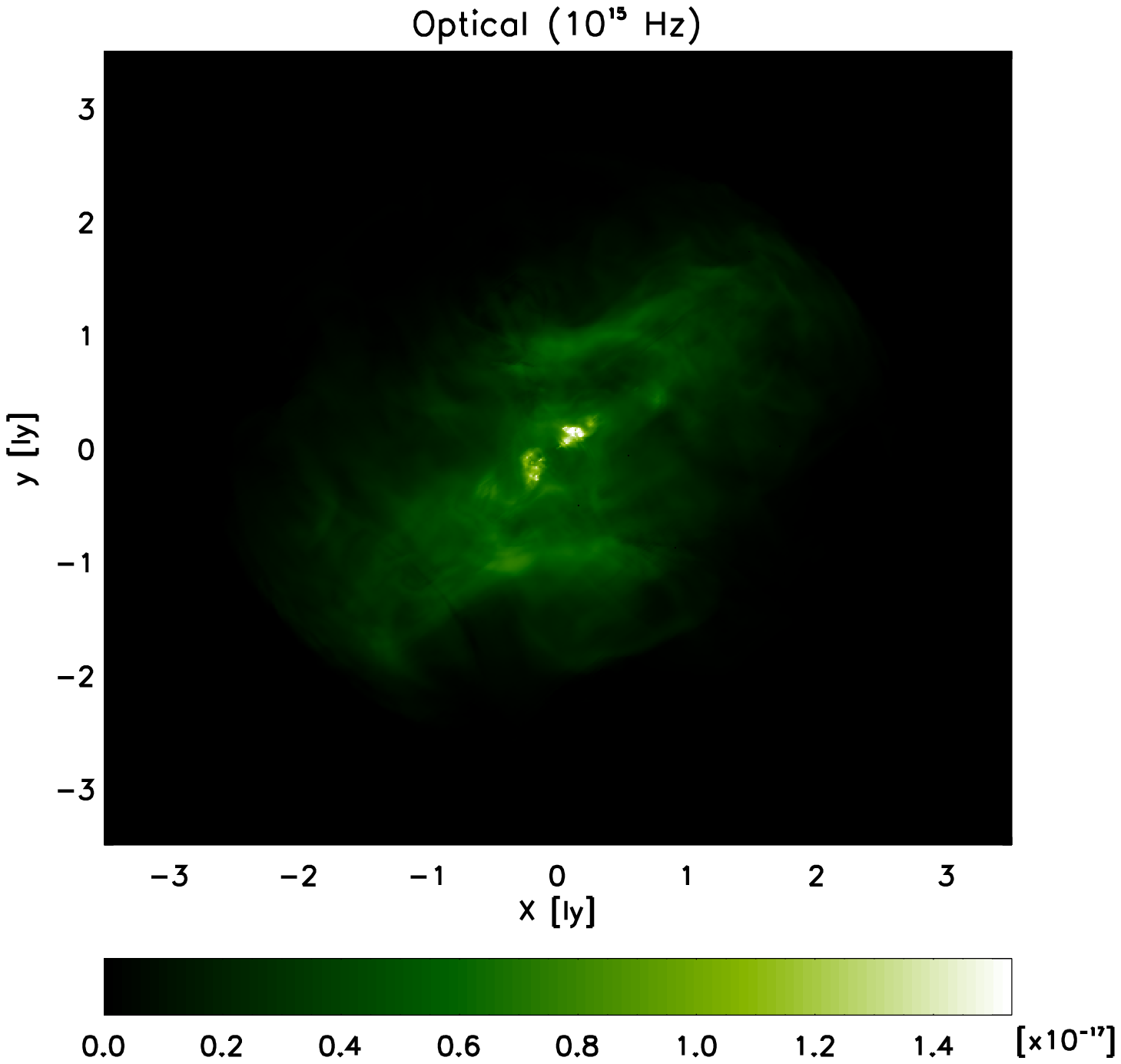}\\
	\includegraphics[scale=0.45]{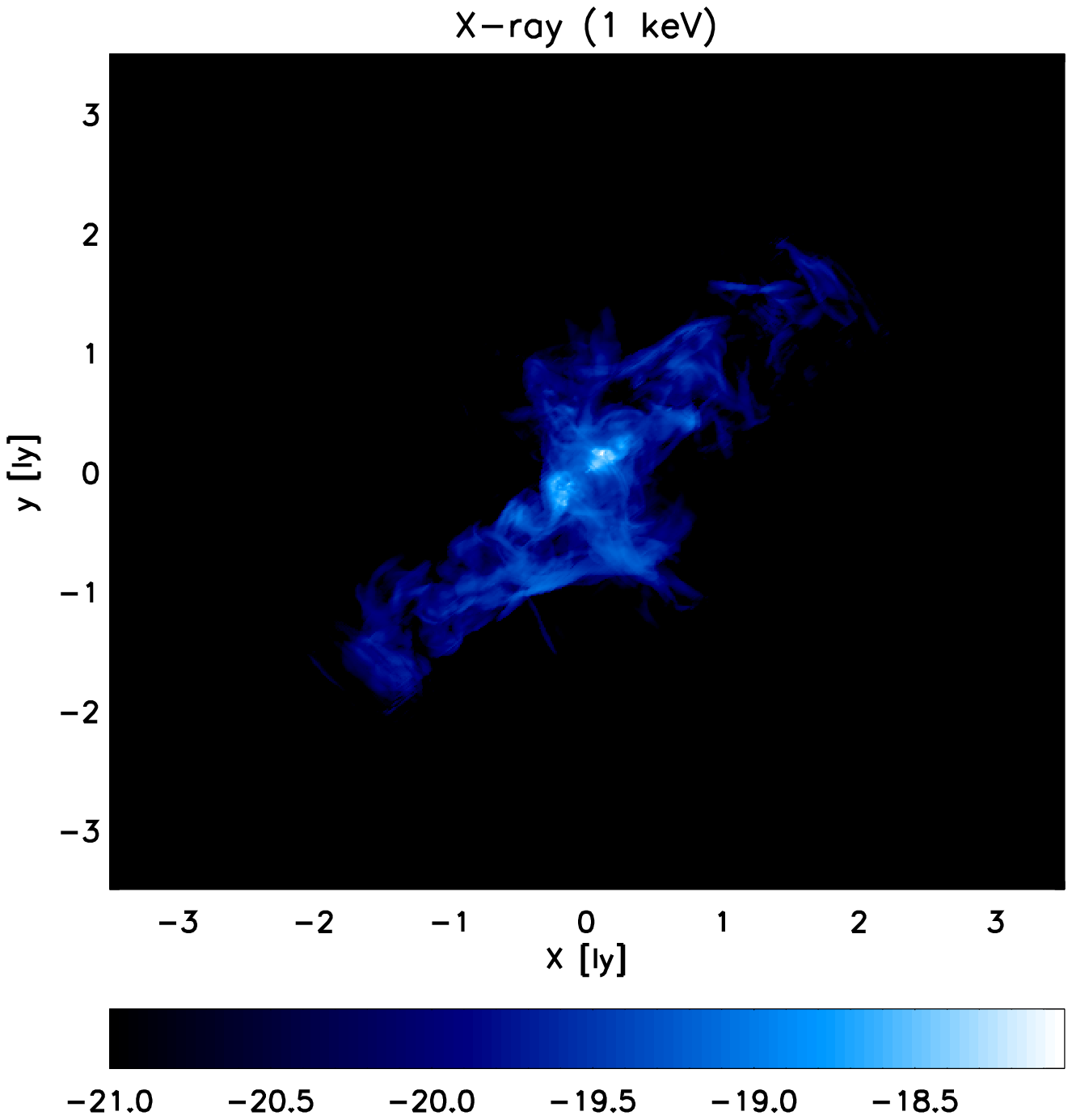}\\	
	\caption{\textit{Left panel, top row}: Map of the radio surface brightness at 5 GHz ($t=250$ yr, corresponding to the evolution time $t_\mathrm{eq}\simeq 500$ yr). 
	Intensity is indicated by the color bar, in linear scale. 
	 \textit{Right panel, top row}: Linear map of the optical surface brightness at 10$^{15}$ Hz ($t{sim}=250$ yr, $t_\mathrm{eq}\simeq 500$ yr).
	 \textit{Bottom row}: Map of the emission at X-rays (1 keV). 
	 In all the maps the intensity is expressed in mJy/arcsec$^2$ units and distance from the pulsar is in ly.  
	 } 
\label{fig:emission}
\end{figure}


In Fig.~\ref{fig:dynvarie} 2-D plots of the velocity magnitude and the magnetic pressure at t=250 yr are shown, again in the  ${(0,\, y,\, z)}$ plane. High-velocity fluxes in the polar direction are present, with velocity of the order of $\gtrsim 0.7c$. These are also regions of high magnetic pressure. In this regions jets are launched thanks to the magnetic hoop stresses, which collimate the high-velocity flow escaping from the polar sector of the TS in the polar jets. This behaviour can also be easily recognized in the velocity map, shown in the bottom-right panel of the same figure.


The resulting emission morphology computed in radio, optical and X-ray bands at t=250 yr is shown in Fig.~\ref{fig:emission}. 
As can be seen, the synchrotron burn-off effect is clearly apparent at increasing photon energies, and emission maps are encouraging. 
Polar jets are well formed but the torus is quite under-luminous. Most of the emission comes indeed from the jets bases, where the particle density is higher, while in the torus zone, although the magnetic field is still high, the number of particles is considerably lower (compare with Fig.~\ref{fig:dynvarie}, upper panel).

The inner nebula also shows the expected time variability, at least at radio frequencies (5 GHz). Since the outputs of our 3-D simulation are very large, and their damping is rather time consuming, we currently only save snapshots every 5 years of evolution. This time interval is too large to address the X-ray variability, which occurs on time scales of weeks to months, but is adequate to discuss variability in the radio band, where times scales are found to be of order a few years \citep{Bietenholz:2004}.

\begin{figure}
\centering
	\includegraphics[scale=0.42]{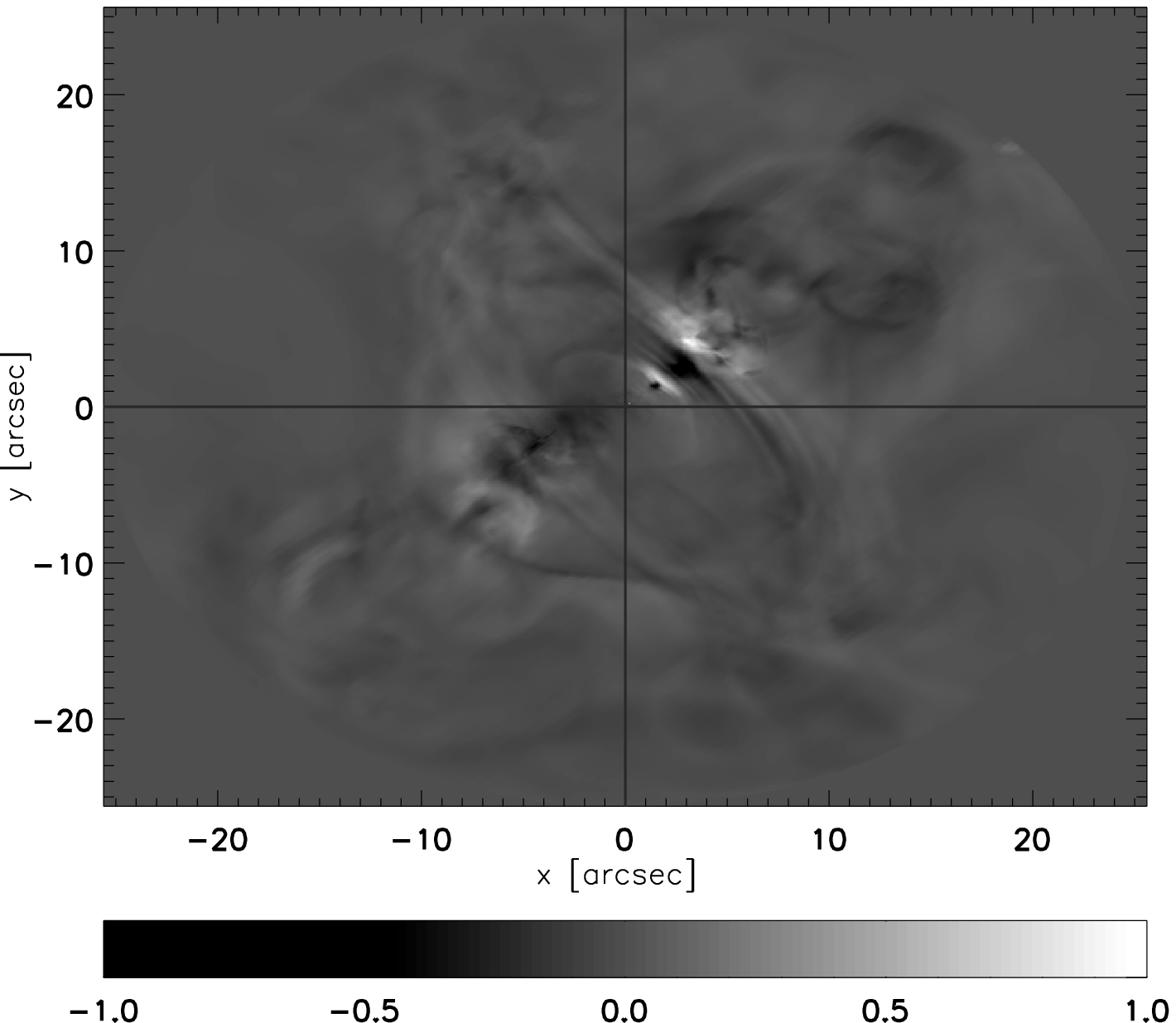}
	\includegraphics[scale=0.42]{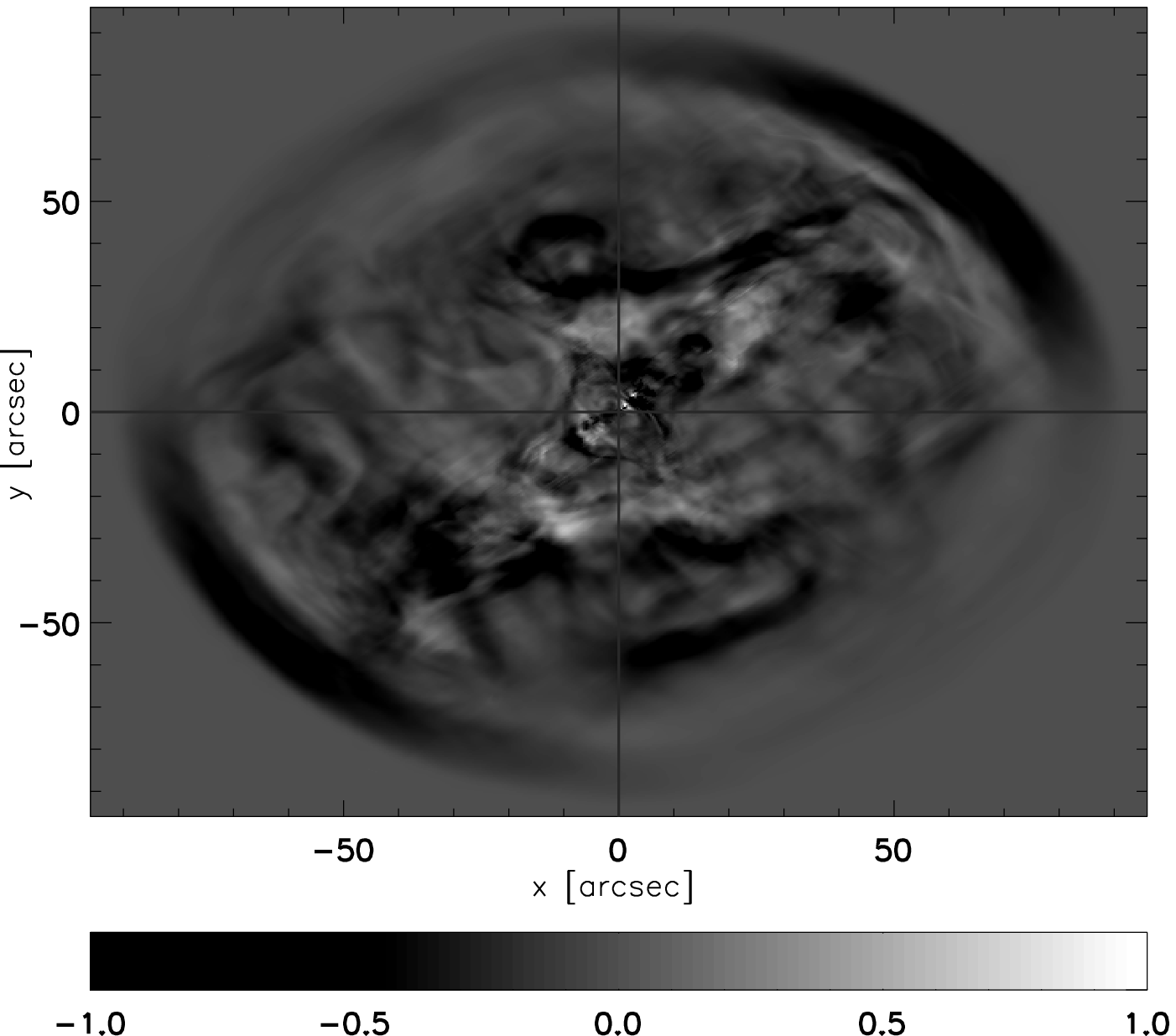}\,
\caption{Map of the wisp-like structures at radio frequencies (5 GHz) in the inner nebula (left panel) and in the whole nebula (right panel). The images are obtained as the difference between two surface brightness maps separated by a time interval of 5 years around t=250 years. The position of the pulsar is identified by the black crosses and intensity is normalized to its maximum.} 
\label{fig:3dwisp}
\end{figure}

In Fig.~\ref{fig:3dwisp} (right panel) wisps are shown as a sequence of consecutive light and dark arcs. They form near the pulsar, the position of which is indicated by the black cross, and then move outward in the bulk of the nebula. As expected, wisps appear to be more prominent in the upper hemisphere of the nebula (with respect to the pulsar equatorial plane) than in the lower one. Properties of wisps are here obtained with the same technique as in the 2-D case, by subtracting emission maps separated by 5 years (in radio the expected periodicity of wisps is of the order of years). 

The results reported in Fig.~\ref{fig:3dwisp}, when compared with those reported in Fig.~\ref{fig:origin2} (from 2-D MHD simulations by \citet{Olmi:2014}), immediately highlight the much greater richness of variable structures that is found in 3-D simulations. A number of non-axisymmetric structures appear on the large scales, and the map shows a much closer resemblance to the time variability data reported in Fig.~2 of \citet{Bietenholz:2004}. However, in the inner nebula the 2-D and 3-D results appear to be rather similar, lending support to the validity of the conclusions that were obtained for that region within the 2-D description.

\section{The \emph{ideal} tearing model for the Crab nebula gamma-ray flares}

As discussed in section~\ref{intro-gamma}, the sudden gamma-ray flares observed in the Crab nebula are most likely due to stochastic reconnection events occurring around the TS region or at the jet base, where the strongly magnetized plasma is supposed to convert its energy into heat and particle acceleration. Unless peculiar magnetic configurations leading to forced (driven) reconnection are considered \citep{Lyutikov:2016}, the most natural process leading to spontaneous reconnection of current sheets is the \emph{tearing} instability. However, even within relativistic resistive MHD or force-free electrodynamics, as appropriate for the hot, highly magnetized plasma in the Crab nebula, the tearing instability is usually regarded as a slow process \citep{Lyubarsky:2005,Komissarov:2007}, precisely as in classical MHD, with a growth time scale which is the geometric mean of the ideal Alfv\'enic one and the much longer diffusion time.

This is actually true if one refers to the current sheet width $a$ as the characteristic length. However, for very thin current sheets, the natural size of the system is rather their length $L$, so that, by rescaling the quantities with the (inverse) aspect ratio $\epsilon = a/L\ll1$, we find a maximum growth rate of the tearing instability modified as:
\begin{equation}
\gamma_\mathrm{max} \, a/c_A \sim S_a^{-1/2} \quad \Rightarrow \quad
 \gamma_\mathrm{max} \, L/c_A \sim S_L^{-1/2} \epsilon^{-3/2},
 \label{growth}
\end{equation}
where $c_A$ is the background Alfv\'en velocity and $S_a$ and $S_L$ are the Lundquist numbers ($S_l=l/c_A\eta$, with $l$ a characteristic length scale and $\eta$ the magnetic diffusivity), referred to $a$ or $L$ respectively ($S_a=\epsilon \, S_L$). Suppose now that $\epsilon\sim S_L^{-\alpha}$. For instance, in the steady-state, incompressible, non-relativistic Sweet-Parker model we have $\alpha=1/2$, and the evolution is then expected to be of  explosive type. Provided $S_L$ is large enough, the current sheet is seen to be rapidly disrupted by the so-called \emph{plasmoid instability} \citep{Loureiro:2007,Bhattacharjee:2009,Samtaney:2009,Uzdensky:2010, Sironi:2016}.
In any realistic environment, where current sheets may form as a result of convective motions, occurring on ideal or Alfv\`en time scales, it is natural to expect their progressive thinning. As a result, the system becomes tearing and plasmoid unstable before the Sweet-Parker configuration is realized. The maximum growth rate $\gamma_\mathrm{max}$ in equation~(\ref{growth}) becomes independent on $S_L$, and we can speak of an \emph{ideal} tearing instability \citep{Pucci:2014,Landi:2015,Tenerani:2015,Del-Zanna:2016}.

The theory of the tearing instability has been recently revisited, for the first time, in the case of resistive relativistic MHD, and both analytical and numerical investigation of the classical and \emph{ideal} tearing have been carried out \citep{Del-Zanna:2016a}. In particular, when current sheets with $\epsilon\sim S_L^{-1/3}$ are considered, the expected instability dispersion relation, independent on the value of $S_L$, is retrieved, and both the linear and nonlinear stages are seen to follow a \emph{universal} law, if quantities are normalized against the relativistic Alfv\'en speed (here $4\pi\to 1$)
\begin{equation}
c_A = \frac{B}{\sqrt{e+p+B^2}}\,c,
\end{equation}
where $B$ is the magnetic strength, $e$ the energy density and $p$ the thermal pressure, for the unperturbed fluid ($c_A\to c$ for magnetically dominated plasmas). 

\begin{figure}
\centering
 	\includegraphics[scale=0.40]{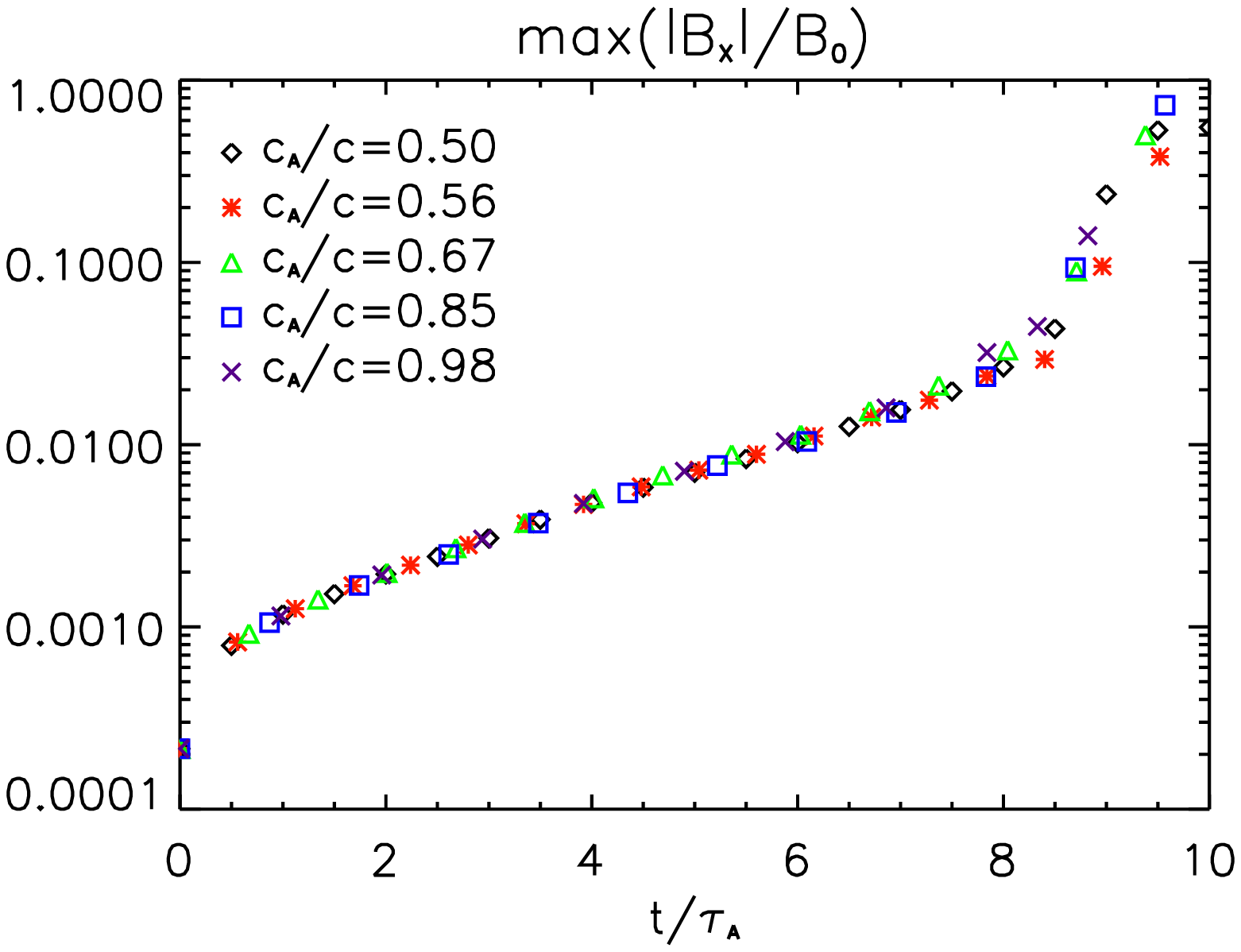} \quad\quad
	\includegraphics[scale=0.34]{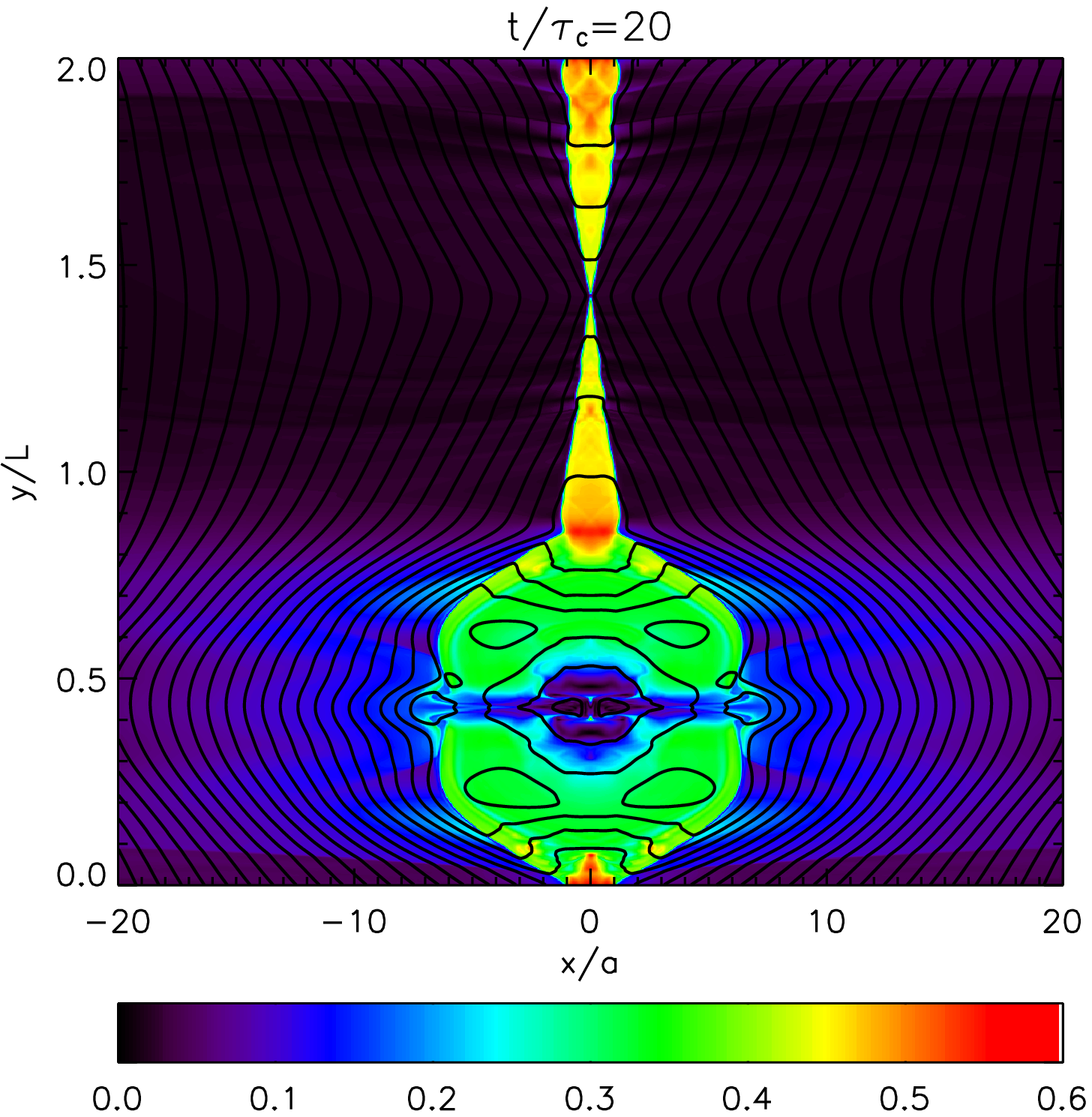}
\caption{Left panel: the growth in time of \emph{r.m.s.} magnetic fluctuations from the initial seed value for different runs, symbols and colors refer to different values of the background Alfv\'en velocity $c_A$. Right panel: color map of the flow velocity and magnetic fieldlines for the case $c_A=0.5c$ at time $t=20\,\tau_c \equiv 10\,\tau_A$.} 
\label{fig:tearing}
\end{figure}

In Fig.~\ref{fig:tearing} (left panel) the averaged magnetic fluctuations across the current sheet (initially set to a seed small value) are shown, as a function of $t/\tau_A \equiv t \, c_A/L$, for five different numerical simulations at increasing $c_A/c$ (from 0.50 up to 0.98): notice the \emph{e}-folding time of order $\tau_A$ (thus an ideal time scale), and a very similar trend even for the nonlinear stage when the plasmoid instability develops, with the merging of magnetic islands and secondary reconnection events.
The final stage of the evolution is, due to the periodic boundary conditions assumed in the sheet direction, a single X-point with Petschek-type funnels containing super-(magneto)sonic jets feeding a large magnetic island. The maximum velocity achieved in jets corresponds to the external Alfv\'en speed, $c_A=0.5\, c$ in the case shown in Fig.~\ref{fig:tearing} (right panel), and we find a quite high reconnection rate which depends only mildly on the Lundquist number, $\mathcal{R}\sim \ln (S_L)^{-1}$, as predicted in the relativistic case for magnetically dominated plasmas \citep{Lyubarsky:2005}.

The result that the \emph{ideal} tearing instability develops within a few light-crossing times of the macroscopic length $L$, independently on microphysics, is of course very important in our attempt to model the gamma-ray flares of the Crab nebula. The measured time scale $\tau$ for growth in the light curve is of approximately one day, so we need current sheets to develop on a scale $L\sim \tau c \sim 10^{15}$~cm. This length is $\sim 1\%$ of the TS radius, compatible with the size of the small-scale variable features observed in the inner regions by HST and Chandra, like the \emph{knot} or the \emph{anvil}, but also with the base of the polar jets, whose kinks are likely to produce sheared magnetic regions. We cannot certainly determine the precise location for reconnection in the Crab nebula from this highly idealized model, but we have demonstrated that the reconnection process is able to occur on ideal time scales, independently on $S_L$ and without invoking kinetic mechanisms, provided that during the current sheet dynamical evolution (thinning) the critical threshold of $\epsilon\sim (S_L)^{-1/3}$ can be reached.

\section{Summary and conclusions}

In the last decade, MHD numerical modelling has been widely recognized
as a powerful tool for the investigation of the physics of the relativistic magnetized plasma in PWNe.

While many open questions still remain, the advent of relativistic MHD simulations has allowed us to answer a number of questions, and has opened a promising way to answer more. 
The initially puzzling jet-torus structures observed in the X-ray emission of a number of nebulae have found a straightforward interpretation as the result of the pulsar wind structure and in particular of the anisotropy of its energy flux. The striking agreement between the simulated and observed high-energy morphology of the 
Crab nebula, which is the PWNe prototype, suggests indeed that the flow structure in the inner regions must be very close to what 2-D MHD simulations predict \citep{Del-Zanna:2006}. 

The open problems concern the exact origin of the low-energy emitting particles, that can be either continuously injected in the nebula as part of the pulsar wind or a relic population, or finally the result of acceleration somewhere else.

This problem was approached by comparing observations with simulations done under different hypotheses for the radio particles origin.
The main conclusion was that the global radio emission is basically insensitive to the spatial distribution of the particles, as long as they are not pure relics. The latter is the only case that can be easily excluded, not only the surface brightness maps are almost identical, but also the variability of the inner regions (radio wisps) is very similar too. This is a result of the fact that within the MHD framework wisps naturally arise from the properties of the flow.
This result should be validated in the future also in three dimensions, since the stronger mixing of the magnetic field lines could in principle allow for the relic scenario to be reconsidered.

It is a matter of fact that wisps are observed at radio, optical and X-ray frequencies, and that they are not coincident at the different wavelengths, with differences measured both in terms of spatial locations and of outward velocities \citep{Bietenholz:2004,Schweizer:2013}. 
As already discussed, within the framework of MHD, the only way to account for the different properties of the wisps at different frequencies is to assume some differences in the spatial distribution of the particles that are responsible for the emission at the different wavelengths. An obvious difference among particles of different energies (and hence emitting in different frequency bands) is the role of energy losses. However, losses are not very important in the innermost regions of the nebula from which wisps are observed. 
Any other difference in the spatial distribution of the particles of different energies suggests differences in the acceleration sites. In other words, if wisps are different at radio and X-ray frequencies, the particles responsible for radio emission must have had a different history than the particles responsible for X-ray emission, and the simplest explanation is that they were accelerated in different locations.

This possibility has been studied by performing 2-D MHD simulations which assume a non-uniform particle acceleration at the TS front. The TS was divided in complementary regions (an equatorial band and a polar cone, with varying angular extent) and particles of different energies were injected in non-coincident ones.
The conclusion is that X-ray wisps are best reproduced if injection in a narrow belt around the pulsar rotational equator is considered. The equatorial region is presumably where the flow magnetization is lower, if effective magnetic dissipation takes place in the striped wind. Then the mechanism responsible for the acceleration of those highest-energy particles could be Fermi I, which has been proven to be very effective at relativistic shocks of low enough magnetization \citep{Spitkovsky:2008, Sironi:2011}, and which naturally provides a particle spectral index close to that inferred from X-ray observations.

Radio emission properties are again much more complicated to constrain: the only scenario that radio observations directly exclude is one in which low-energy particles are injected in a narrow polar cone, since almost no wisps are found.
But emission properties in the case of uniform injection, or injection in a wide polar or equatorial band, are basically indistinguishable.


Since 2010, when the Crab gamma-ray flares were revealed for the firs time \citep{Tavani:2011}, many attempt to explain this explosive events were done. Up to now the most plausible explanation seems to be violent stochastic reconnection events occurring in the TS region or at the jet base. 

This can be obtained  thanks to either collapsing magnetic flux bundles or other peculiar configurations, leading to explosive reconnection and particle acceleration \citep{Lyutikov:2016}, or via the spontaneous reconnection in thinning current sheets, the so-called \emph{ideal} tearing instability, which is also proved to develop on extremely rapid time scales, basically the light-crossing time of the sheet length. \citet{Del-Zanna:2016a} have shown that this mechanism can account for the correct time variation scaling observed in the light curves of gamma-ray flares by considering current sheets which develop on scales of $\sim 1\%$ of the TS radius, which are compatible with the scales of the bright variable features observed in the inner nebula and at the base of the polar jets.

Although axisymmetric 2-D models have allowed us to investigate many of the problems connected to the physics at work in the inner nebula, they were not able to reproduce the bulk properties of the nebula at larger scales. 
In fact they point to values of the wind magnetization of the order of $10^{-2}$, which are incompatible with the integrated emission spectrum, since they lead to a nebular magnetic field which is too low to account for the whole emission. As already highlighted by \citet{Volpi:2008}, reproducing the synchrotron part of the spectrum requires
conversion of the flow energy flux into accelerated particles with efficiencies larger than one, over-predicting the IC emission. 
Moreover the suppression of the third dimension also produces an artificial pile up of the magnetic field morphology, which prevents to compute reliable emission properties of the outer nebula, and also to reproduce the kinking jets.

The first 3-D modelling seems to provide a solution to this problem \citep{Porth:2013, Porth:2014}. These models show that kink instabilities, which are artificially suppressed in two dimensions, efficiently reduce the hoop stresses that force the magnetization to have such low values, allowing reaching values of the order of unity. 
Recent works about current driven instabilities in three dimensions \citep{Mizuno:2011,Mignone:2013}
confirm in fact the formation of kinks in jets, even if the flow arising from the termination
shock and the hoop stresses in the nebula itself are not taken into account.

These first 3-D simulations seem to provide a reasonable solution to the long-standing $\sigma-$paradox and to account for the formation of a PWN with features similar to those observed in the Crab nebula, although the jets are very weak and almost under-luminous. Moreover they only cover a very short time ($\sim 70$ yr) after the SN explosion, and at the end of the simulation the self-similar expansion phase has not been reached yet.
Here we have presented our preliminary results of the first long-term 3-D simulations of the Crab nebula.
We show that, in just $\sim200$ yrs the nebula reaches the self-similar phase, and thus the inferred information should be directly compared with observed properties.

As also shown by \citet{Porth:2014}, the onset of the kink instability leads the initial toroidal field to be strongly modified during the evolution of the system, with the poloidal component which suddenly develops and becomes as important as the toroidal one near jets.
As soon as the self-similar phase has been reached the strong magnetic dissipation, which lowers the initial
field to average values of approximately 100 $\mu$G, slows down.
Extrapolation of this trend is promising to lead the field to reach the equipartition value at the final stage of the evolution.
Nevertheless the volume averaged field is still lower than expected (by a factor of $\sim2$), which may indicate that the dissipation is still too high in the first stages of the evolution.

Our simulations show that the dynamics of the inner nebula is in good agreement with what was found in 2-D simulations, suggesting that 2-D models are absolutely robust in describing the physics of the inner regions. This is obviously not true anymore when one moves from the pulsar vicinity to the outer nebula, where the effects of the poloidal component of the field become important.

We compute emission properties at radio, optical and X-ray frequencies. The synchrotron burn-off effect is seen through the decrease of the size of the emitting area with increasing observation frequency. However the brightness contrast of the torus is rather low when compared with observations, while the brightest features appears to lie at the bases of the jets, where the particle pressure is higher.

We also compute maps of the time variability at radio frequencies for both the wisps region (the inner nebula) and for the entire nebula, obtaining very encouraging results. The simulated radio wisps confirm again the similarities between results from 2-D and 3-D simulations in reproducing the properties of the inner nebula. In the outer regions, on the other hand, the 3-D results resemble the observations much more closely than those in two dimensions, with the appearance of moving structures near the polar jets and a number of additional arc-like features in the body of the nebula.

As soon as the simulation will have reached an evolved enough stage, we plan to address the issue of the integrated synchrotron and IC spectrum, and gather insight on the actual number of particles in the wind.
The goal of constraining the pulsar wind physics, by finding a suitable set of parameters that allows to reproduce both the morphology and multi-wavelength spectrum of the Crab Nebula, finally seems within reach.

\section*{}
B.O. and L.D.Z acknowledge support from the INFN - TEONGRAV initiative (local PI: L.D.Z) and from the University of Florence grant \enquote{Fisica dei plasmi relativistici: teoria e applicazioni moderne}. N.B. acknowledges support from the NSMAG (EU FP7 - CIG) project (PI: N.B.).

	\footnotesize{
	\bibliographystyle{jpp}
	\bibliography{olmi}
	}

\end{document}